\documentclass[twocolumn,tighten,times]{aastex631}

\newcommand{\um}{$\mu$m}

\accepted{April 7, 2023; ApJ}

\shorttitle{Thermal polarized emission of Radio-Loud and Radio-Quiet AGN}
\shortauthors{Lopez-Rodriguez, E.}
\graphicspath{{./}{figures/}}

\begin{document}

\title{On the origin of radio-loudness in active galactic nuclei using far-infrared polarimetric observations}

\correspondingauthor{Enrique Lopez-Rodriguez}
\email{elopezrodriguez@stanford.edu}

\author[0000-0001-5357-6538]{Enrique Lopez-Rodriguez}
\affiliation{Kavli Institute for Particle Astrophysics \& Cosmology (KIPAC), Stanford University, Stanford, CA 94305, USA}

\author{Makoto Kishimoto}
\affiliation{Department of Physics, Kyoto Sangyo University, Motoyama, Kamigamo, Kita-ku, Kyoto, 603-8555, Japan}

\author{Robert Antonucci}
\affiliation{Department of Physics, University of California in Santa Barbara, Broida Hall, Santa Barbara, CA 93109, USA}

\author[0000-0003-0936-8488]{Mitchell C. Begelman}
\affiliation{JILA, University of Colorado and National Institute of Standards and Technology, 440 UCB, Boulder, CO 80309-0440, USA}
\affiliation{Department of Astrophysical and Planetary Sciences, 391 UCB, Boulder, CO 80309-0391, USA}

\author[0000-0001-9011-0737]{Noemie Globus}
\affiliation{Department of Astronomy and Astrophysics, University of California, Santa Cruz, CA 95064, USA}
\affiliation{Center for Computational Astrophysics, Flatiron Institute, Simons Foundation, New-York, NY10003, USA}

\author[0000-0002-1854-5506]{Roger Blandford}
\affiliation{Department of Physics, Stanford University, Stanford, California 94305, USA}
\affiliation{Kavli Institute for Particle Astrophysics \& Cosmology (KIPAC), Stanford University, Stanford, CA 94305, USA}

\begin{abstract}

The dichotomy between radio-loud (RL) and radio-quiet (RQ) active galactic nuclei (AGN) is thought to be intrinsically related to radio jet production. This difference may be explained by the presence of a strong magnetic field (B-field) that enhances, or is the cause of, the accretion activity and the jet power. Here, we report the first evidence of an intrinsic difference in the polarized dust emission cores of four RL and five RQ obscured AGN using $89$ \um\ polarization with HAWC+/SOFIA. We find that the thermal polarized emission increases with the nuclear radio-loudness, $R_{20} = L_{\rm 5GHz}/ L_{\rm 20\mu m}$. The dust emission cores of RL AGN are measured to be polarized, $\sim5-11$\%, while RQ AGN are unpolarized, $<1\%$. For RQ AGN, our results are consistent with the observed region being filled with an unmagnetized or highly turbulent disk and/or expanding outflow at scales of $5-130$ pc from the AGN. For RL AGN, the measured $89$ \um\ polarization arises primarily from magnetically aligned dust grains associated with a $5-130$ pc-scale dusty obscuring structure with a toroidal B-field orientation highly offset, $65\pm22^{\circ}$, with respect to the jet axis. Our results indicate that the size and strength of the B-fields surrounding the AGN are intrinsically related to the strength of the jet power---the stronger the jet power is, the larger and stronger the toroidal B-field is. The detection of a $\le130$ pc-scale ordered toroidal B-field suggests that a) the infalling gas that fuels RL AGN is magnetized, b) there is a magnetohydrodynamic wind that collimates the jet, and/or c) the jet is able to magnetize its surroundings. 

\end{abstract}




\section{Introduction} \label{sec:INT}

\subsection{Origin of the radio-loudness in active galaxies}
Active galactic nuclei (AGN) have been historically divided into radio-loud (RL) and radio-quiet (RQ) based on the relative ratio between their radio and optical luminosities \citep[e.g.,][]{Visnovsky1992,Stocke1992,Kellermann1994}. The most common definition is given by $R=L_{5GHz}/L_{\rm B}$, where $L_{5GHz}$ is the luminosity at $5$ GHz, and $L_{\rm B}$ is the luminosity in the B-band. The typical cut-off is at $R=10$. For an optically selected sample, quasars have been found to be mostly RQ where $\sim10-20$\% are RL \citep[e.g.,][]{Visnovsky1992}. These studies typically use the integrated luminosities of the entire galaxy. Although this may be valid for quasars, where the jet outshines the host galaxy, it may not be fully valid for Seyfert galaxies. In addition, beaming effects can also affect both the optical and the radio radiation. Using a morphological classification, RL AGN are mainly characterized by large-scale ($10$s of kpc) radio jets and lobes with their kinetic powers comprising a significant fraction of the total bolometric luminosity. Small-scale jets ($100$s of pc) with thermal emission by dust and/or accretion disk along the equatorial plane are the dominant structures of RQ AGN. Aside from the synchrotron component arising from the radio jet, the total flux spectral energy distributions (SEDs) from the infrared (IR) to the X-ray of RQ and RL AGN are virtually identical \citep[e.g.,][for a review]{Sanders1989,Elvis1994,Hickox2018}. Thus, the RQ and RL AGN dichotomy is thought to be intrinsically related to the radio jet production and/or stage of activity. 

Advected magnetic fields (B-fields) are strongly implicated in all models that explain the launch and collimation of relativistic jets \citep[][for a review]{Blandford2019}. Within the central pc, the presence of a radio source at the core and collimated jets imply a manifestation of the strong accretion flow onto the supermassive black hole (SMBH). Accretion flow towards SMBH can be supported by small-scale ($<<$ pc) B-fields generated by dynamo action and B-field lines stretching in a turbulent shear flow \citep[e.g.,][]{Brandenburg1995,Sano2004,Beckwith2011,Simon2012}. The transport of angular momentum is suggested to be operated by magnetorotational instability \citep[MRI;][]{Balbus1991,Balbus1998}, which gives rise a sustained large-scale toroidal B-field and magnetohydrodynamical (MHD) turbulence. The MHD fluctuations within the disk produce a viscous stress responsible for the outward angular momentum transport driving gas accretion. 

There are several physical mechanisms sustaining a large-scale B-field in AGN. Let us first consider the infall of gas to form a disk. The natural outer radius for a disk is given by the radius of influence $r_{\rm inf}=GM_{\rm BH}/ \sigma^{2} \sim10^6$ times the gravitational radius, where the black hole mass is $M_{\rm BH}$ and the central velocity dispersion is $\sigma$. Within this radius, gravity is dominated by the black hole, and a thin disk is supposed to form. The thin disk allows gas to move inwards under magnetic and gravitational torques. However, thin accretion disks have been found to be inefficient at depositing magnetic flux close to the SMBH. Thin disks make most of the AGN to be RQ. In addition, thin accretion disks do not generate long-term large-scale poloidal B-fields that connect the accretion disk and the SMBH in RL AGN \citep{Lubow1994}. Alternatively, if the infalling gas has a specific angular momentum $> GM_{\rm BH} / \sigma$, it is likely to form a disk before it reaches $r_{\rm inf}$. The angular velocity will vary more slowly with increasing radius. Under this scenario, a toroidal B-field within a disk beyond the radius of influence could be generated. 

Hot accretion in thick disks has been proposed to explain the accretion onto SMBH \citep[e.g.,][]{Ichimaru1977,Rees1982,Lubow1994,GO2012}. Geometrically thick accretion disks (i.e., magnetically elevated disks) are supported vertically against the vertical component of gravity by a strong toroidal B-field generated by the accretion disk dynamo \citep{Begelman2017}. These accretion disks are thicker and less dense than standard disks, with stratified layers along the vertical axis. The magnetic pressure elevates most of the accretion flow to large heights from the midplane \citep{Begelman2017,Zhu2018,Mishra2020}. The midplane consists of a thin layer with most of the disk mass but with little contribution to the accretion rate. Under this scheme, the elevated layers can intercept a fraction of the incident radiation from the AGN and regulate the density and motion of the gas. These models may imply that there should be a steady loss of flux through Parker instabilities. These magnetically elevated disks may explain the accretion flow from pc to sub-pc scales, broad line emission, and the relation between star formation and accretion in AGN \citep{Begelman2017}. 

Another possibility is that an extended disk drives a large, unipolar MHD wind, which carries off mass, angular momentum, and energy from the disk \citep{BG2022}. The B-field in such a wind will become increasingly toroidal and perhaps unstable. The infall is predominantly happening in the midplane of the disk at large scales ($>$ pc), while an MHD wind transports outward the energy from the SMBH within the Bondi radius. The MHD wind supports the jet’s collimation \citep[example in figure 1 a-b by][]{BG2022}.

The fundamental difference between RL AGN and their RQ AGN siblings may indeed be due to the presence or absence of a strong and ordered toroidal B-field. This ordered toroidal B-field enables the collimation of the jet and an effective transfer of angular momentum and energy outwards from the SMBH. As the jet power depends on and/or generates the B-fields in the surrounding environment of the AGN, one may expect that the stronger the jet power is, the larger and stronger the ordered toroidal B-field is. This scenario invites the question of the origin and transportation of the B-fields from $1-100$ pc scales to sub-pc scales, which ultimately connects the host galaxy dynamics with the AGN. Within this framework, this manuscript aims to measure the signature of a magnetized flow at scales of $<100$ pc in a far-infrared flux-limited sample of RQ and RL AGN.

\subsection{Infrared observational signatures of B-fields in AGN}\label{subsec:Intro_IRPol}

The unified model of AGN \citep{Antonucci1993,UP1995} posits that a dusty and molecular structure (the `torus’\footnote{Note that observations only show that a thick equatorial band is opaque to optical radiation, as seen from the nucleus. `Torus' is a shorthand, but it refers only to the shadowing properties of the obscuration, with no implications for the outer boundary, the dynamics, the morphology, or any other physical property.})  may surround some AGN that absorbs radiation at all wavelengths and re-emits it at infrared (IR) wavelengths. The postulated torus explains the large varsiety of AGN types from an orientation point of view.  At pc-scales, the gas becomes dusty when the temperature is $\le1300$ K (sublimation temperature of graphite), representing the inner wall of a dusty environment surrounding the active nuclei. Seyfert galaxies have shown to have a dusty and molecular, geometrically thick obscuring disk along the AGN equatorial plane up to a median radius of $\sim42$ pc with a median molecular gas mass of $\sim6\times10^{5}$ M$_{\odot}$ \citep{GB2021}. This disk is rotating and resembles the inflowing gas in an accretion disk toward the SMBH with an outflowing component above and below the disk. These results were obtained using continuum emission at $870$ \um\ and CO(3-2) and HCO$^{+}$(4-3) molecular emission lines with the Atacama Large (sub-)Millimeter Array (ALMA). Most of the mid-IR (MIR) emission has been found to arise from the thermal emission of dust in pc-scales outflows above and below the equatorial axis of the torus \citep[e.g.,][]{Hoenig2012,Hoenig2013,LG2016,GR2021}. Thus, the dusty and molecular torus comprises a region where outflows and inflows can coexist \citep[e.g.,][]{Honig2019}. 

\begin{deluxetable*}{lcccp{1.8cm}ccccc}[ht!]
\tablecaption{Summary of new OTFMAP polarimetric observations with HAWC+. a) Object name. b) Central wavelength of the band used for observations (\um). c) Observation date (YYYYMMDD). d) Flight ID. e) Sea-level altitude during the observations (ft.). f) Speed of the scan (\arcsec/sec.). g) Amplitudes in elevation (EL) and cross-elevation (XEL) of the scan (\arcsec). h) Time per scan (s). i) Number of observation sets used (and rejected). j) Total on-source observation time (s).
\label{tab:table1}}
\tablecolumns{7}
\tablewidth{0pt}
\tablehead{\colhead{Object}& \colhead{Band} & \colhead{Date}	&	\colhead{Flight}	&	\colhead{Altitude}	&
\colhead{Scan Rate} &  \colhead{Scan Amplitude } & \colhead{Scan Duration} & \colhead{\#Sets (bad)} & \colhead{t} \\ 
&	\colhead{(\um)}  & \colhead{(YYYYMMDD)}	&		&	\colhead{(ft)}	&
\colhead{(\arcsec/sec)} &  \colhead{ (EL $\times$ XEL; \arcsec)} & \colhead{(s)}  & & \colhead{(s)} \\
\colhead{(a)} & \colhead{(b)} & \colhead{(c)} & \colhead{(d)} & \colhead{(e)} & \colhead{(f)} & \colhead{(g)} & \colhead{(h)} & \colhead{(i)} &\colhead{(j)}
}
\startdata
Centaurus~A	&	53	&	20190717		&	F597	&	39000-40000	&	100	&		60$\times$40	&	100	&	4	&	1600	\\	
Mrk~231		&	89	&	20190918		&	F611	&	40000	&	100	&		50 $\times$ 50	& 100	&	10(5) 	& 2000 \\
NGC~1275	&	89	&	20190905		&	F606	&	40000-42000	&	100	&		50 $\times$ 50	&	120	&	11(1)	&	4800\\
			&		&	20190907		&	F607	&	40000-41000	&	100	&		50 $\times$ 50	&	120	&	7(1)	&	2880	\\
NGC~4151	&	89	&	20200128		&	F654	&	41000	&	100, 200	&		100$\times$100, 60$\times$60	&	123	&	8(2)	&	3444	\\
			&		&	20200129		&	F655	&	39000	&	200	&		100$\times$100	&	123	&	5	&	2460	\\	
NGC~5506	&	89	&	20200131		&	F657	&	43000-44000	&	100	&		50 $\times$50	&	123	&	6	&	2952	\\
PKS~1345+125	&	89	&	20200129	&	F655	&	42000-43000	&	100	&		80$\times$80	&	112	&	15	&	6720	\\
\enddata
\end{deluxetable*}

The magnetic signature in the dusty environment surrounding AGN can be measured by means of magnetically aligned dust grains. The elongated and paramagnetic dust grains can be aligned by the presence of B-fields described by theories of radiative torques (RATs) and also by intense radiation fields or outflowing media \citep[e.g.,][]{Andersson2015,HL2016}. Specifically, dust grains absorb more radiation along one of their axes, making them spin up along their greatest moment of inertia (i.e., short axis). Then, the dust grains acquire a magnetic moment, making them precess along the orientation of the local B-field. The final configuration is one where the short axis of the dust grains is parallel to the local B-field.  As radiation propagates through these magnetically aligned dust grains, preferential extinction of radiation along one plane leads to a measurable polarization in the transmission of this radiation, a process called dichroic absorption. For dichroic absorption, the observed position angle (PA) of polarization traces the orientation of the B-field. The same magnetically aligned dust grains re-emit radiation at MIR and far-IR (FIR), $12-500$ \um, wavelengths preferentially along their major axis, a process called dichroic emission. For dichroic emission, the observed PA of polarization is perpendicular to the local B-field orientation \citep[e.g.,][]{Hildebrand1988,Aitken2004}.

Recent works have shown that there may be an intrinsic difference in the nuclear thermal polarization between RQ and RL AGN. The general trends are: 1) some RL AGN have high intrinsically polarized ($\sim10$\%) cores at $2-90$ \um\ dominated by magnetically aligned dust grains in the dusty torus at scales of $1-100$ pc \citep{ELR2013,ELR2015,ELR2018b,ELR2021}, and 2) RQ AGN have low polarized ($\le1$\%) cores within the $10-90$ \um\ wavelength range \citep{ELR2016,ELR2018c,ELR2020}, with their cores showing a variety of physical mechanisms (i.e., dust scattering, self-absorbing dust emission) at scales of $\sim10$s pc. Thus, the FIR polarimetric results may indicate that the dusty torus represents an interface of a magnetized flow at $1-100$ pc scales and a potential clue for the difference between RQ and RL AGN. Note that some RL AGN also have highly polarized cores at $2$ \um\ \citep[i.e.][]{Capetti2000,Tadhunter2000,Ramirez2014}, where the nuclear polarization arises mainly from scattering off dust located in the surrounding regions of the AGN. These studies analyzed the unresolved nuclear polarization using a single wavelength, which may be affected by several competing polarization mechanisms (i.e., scattering, dichroism, and synchrotron). A multi-wavelength polarimetric analysis is required to disentangle the physical structure associated with the dominant polarization mechanism across the SED \citep{ELR2018c,Marin2018,Marin2020}. 

In this manuscript, we measure the $53-214$ \um~thermal polarized emission of magnetically aligned dust grains in the dusty environment surrounding RL and RQ AGN. The results of this work are discussed in the context of a magnetized flow connecting the AGN with the host galaxy and the SMBH. We describe the specifics of our observations in Section \ref{sec:OBS}. Section \ref{sec:Analysis} shows the analysis of the total and polarized SEDs of our AGN sample. The radio-loudness classification and comparison with the observed thermal polarized emission are presented in Section \ref{sec:R}. Our discussions are described in Section \ref{sec:DIS}, and our main conclusions are summarized in Section \ref{sec:CON}. In this paper, we take $H_{0} = 67.8$ km s$^{-1}$ Mpc$^{-1}$, $\Omega_{\rm M} = 0.308$, $\Omega_{\rm V} = 0.692$.


\section{FIR polarimetric observations} \label{sec:OBS}

\begin{deluxetable*}{lcccccccccccc}
\tablecaption{Object sample with nuclear radio and MIR properties. (a) Object name. (b) Redshift. (c) Luminosity distance in Mpc. (d) Spatial scale in pc/". (e) Nuclear fluxes at $5$ GHz in Jy. (f) Luminosity at $5$ GHz in erg s$^{-1}$ using Eq. \ref{eq:L5}. (g) Nuclear fluxes at $20$ \um\ in Jy. (h) Luminosity at $20$ \um\ in erg s$^{-1}$ using Eq. \ref{eq:L20}. (i) Radio-loudness in log-scale using Eq. \ref{eq:R20}.
\label{tab:table2}}
\tablewidth{0pt}
\tablehead{\colhead{Object}	&	\colhead{z}	&	\colhead{$D_{\rm L}^{(1)}$}	&	\colhead{Scale}		&	\colhead{F$_{\rm 5GHz}^{(2)}$} & \colhead{$L_{\rm 5GHz}$}	&	\colhead{F$_{20\mu m}^{\dagger}$}	&	\colhead{$L_{20\mu m}$}	& \colhead{$\log R_{20}$}	\\ 
						&	&	\colhead{(Mpc)}	&	\colhead{(pc/")}		& \colhead{(Jy)}	&	\colhead{(erg s$^{-1}$)} &	\colhead{(Jy)}	&	\colhead{(erg s$^{-1}$)}	\\
						&	&		&	&		&  	(Eq. \ref{eq:L5})	&		& (Eq. \ref{eq:L20})	& (Eq. \ref{eq:R20})	\\
\colhead{(a)} & \colhead{(b)} & \colhead{(c)} & \colhead{(d)} & \colhead{(e)} & \colhead{(f)} & \colhead{(g)} & \colhead{(h)} & \colhead{(i)}  
}
\startdata
Centaurus~A		& 0.00183	& 3.42	&	16		& 5.12$\pm$0.25	&	38.55	&	$2.02\pm0.09$	&	41.63	& 0.40		\\
Circinus			& 0.00145	& 4.21	&	19		& 0.115$\pm$0.012	&	37.08	&	$13.47\pm0.22$	&	42.63	& -2.07		 \\
Cygnus~A 		& 0.05607	& 255	&	1110 	& 3.26$\pm$0.16	&	42.06	& 	$0.048\pm0.011$	& 44.03$^{\dagger}$	&1.51	 \\
Mrk~231 			& 0.04217	& 195	&	868		& 0.155$\pm$0.008	&	40.51	&	$1.47\pm0.15^{\dagger\dagger}$	& 45.28$^{\dagger}$	&	-1.29			 \\
NGC~1068 		& 0.00379	& 14.4	&	70		& 0.085$\pm$0.004	&	38.02	&	$16.61\pm0.69$	& 43.86	&	-2.36	 \\
NGC~1275		& 0.0175	& 62.5	&	357		& 28.17$\pm$1.69	&	41.80	&	$0.89\pm0.05$	& 44.09$^{\dagger}$	&	1.19		\\
NGC~4151		& 0.0033	& 18.3	&	88		& 0.125$\pm$0.008	&	38.39	& 	$1.13\pm0.03$	& 43.13$^{\dagger}$	&	-1.26			\\
NGC~5506		& 0.0061	& 28.7	&	128		& 0.114$\pm$0.006	&	38.74	& 	$1.80\pm0.3$	&	43.42	&	-1.20		 \\
PKS~1345+125 	& 0.1202	& 591	&	2274		& 3.1$\pm$0.3		&	42.72	& 	$0.41\pm0.02^{\dagger\dagger\dagger}$	&	45.36		&	0.84			\\
\enddata
\tablenotetext{^{\dagger}}{The fluxes were multiplied by a factor of two to scale the $11.6$ \um\ fluxes to $20$ \um\ fluxes. Except when indicated, $11.6$ \um~fluxes were obtained from \citet{Asmus2014}.}
\tablenotetext{^{\dagger\dagger}}{Photometric data at $11.6$ \um\ using the Gran Telescopio CANARIAS \citep{ELR2017}.}
\tablenotetext{^{\dagger\dagger\dagger}}{Photometric data at $20$ \um\ using \textit{Spitzer} \citep{Veilleux2009}.}
\tablenotetext{}{References of the luminosity distance and flux at $5$ GHz. Centaurus~A: $^{1}$\citet{Ferrarese2007} $^{2}$\citet{Schreier1981}. Circinus: $^{1}$\citet{Karachentsev2013} $^{2}$\citet{Elmouttie1998}. Cygnus~A: $^{1}$\citet{Stockton1994} $^{2}$\citet{Perley1984}. Mrk~231: $^{1}$\citet{Carilli1998} $^{2}$\citet{Ulvestad1981}. NGC~1068: $^{1}$\citet{BH1997} $^{2}$\citet{WU1983}. NGC~1275: $^{1}$\citet{Tully2013} $^{2}$\citet{Liuzzo2010}. NGC~4151: $^{1}$\citet{Bottinelli1984} $^{2}$\citet{Ulvestad1981}. NGC~5506: $^{1}$\citet{Tully1988} $^{2}$\citet{Ulvestad1984}. PKS~1345+125: $^{1}$\citet{Horiuchi2004} $^{2}$\citet{Horiuchi2004}.}
\end{deluxetable*}

A flux-limited sample of RQ and RL AGN was performed using the High-resolution Airborne Wideband Camera-plus \citep[HAWC+;][]{Vaillancourt2007,Dowell2010,Harper2018} on the $2.7$-m Stratospheric Observatory For Infrared Astronomy (SOFIA) telescope. The observing constraints were a) objects with total intensity $>1$ Jy integrated within the beam size at $89$ \um, and b) $\le2$h on-source time to achieve a signal-to-noise in the polarization fraction $\ge3$ for an assumed polarization of $5$\%. Table \ref{tab:table1} shows the log of observations for the new objects observed as part of a pilot study during Cycle 7 (ID: 07\_0032, PI: Lopez-Rodriguez, E.). HAWC+ polarimetric observations simultaneously measure two orthogonal components of linear polarization arranged in two arrays of $32 \times 40$ pixels each. All objects were observed at $89$ \um~ with a pixel scale of $4$\farcs$02$ pixel$^{-1}$, and beam size (full width at half maximum, FWHM) of $7$\farcs$80$. For bright objects (Centaurus~A, Circinus, Cygnus~A, and NGC~1068), observations also at $53$ and $214$ \um~were performed with pixel scales of $2\farcs55$ and $9\farcs37$ pixel$^{-1}$ and beam sizes of $4\farcs85$ and $18\farcs2$, respectively. All observations were performed using the on-the-fly-map (OTFMAP) polarimetric mode. The OTFMAP polarimetric mode has been successfully applied to galaxies, Centaurus A \citep{ELR2021}, NGC~1097 \citep{ELR2021c}, as well as the filamentary cloud L1495/B211 \citep{Li2021}. The full description of the OTFMAP polarization mode of HAWC+ is described by \citet{ELR2022a}. Here, we present a summary of the OTFMAP polarimetric observations and data reduction.

\begin{deluxetable*}{lcccccccl}[ht!]
\tablecaption{Photometric and polarimetric measurements using HAWC+ observations. (a) Object name. (b) Central wavelength of the HAWC+ band (\um). (c) Angular (\arcsec) and spatial (kpc) size of the aperture used for the photometric and polarimetric analysis. (d) Total intensity in Jy. (e) Polarization fraction in \%. (f) Position angle of polarization of the E-vector in degrees. (g) Polarized flux in mJy. (h) Intrinsic polarization fraction as estimated in Section \ref{subsec:IP}. (i) References of the polarization measurements.
\label{tab:table3}}
\tablecolumns{7}
\tablewidth{0pt}
\tablehead{\colhead{Object}& \colhead{Band} 	&	\colhead{Aperture}		&	 \colhead{$I$}	&	\colhead{$P_{\rm m}$}	&	\colhead{$PA_{\rm E}$}	& \colhead{$PI$}  & \colhead{$P_{\rm int}^{(a)}$} & \colhead{Reference}\\ 
					&	\colhead{(\um)}  &	\colhead{(\arcsec) / (kpc)}	& \colhead{(Jy)}	&	\colhead{(\%)}	&	\colhead{($^{\circ}$)}		&	\colhead{(mJy)}	 & \colhead{(\%)} &	\\
\colhead{(a)} & \colhead{(b)} & \colhead{(c)} & \colhead{(d)} & \colhead{(e)} & \colhead{(f)} & \colhead{(g)} & \colhead{(h)} & \colhead{(i)}
}
\startdata
Centaurus A		&	53	&	5.0/0.08		&	$10.5\pm1.2$		&	$1.9\pm0.6$	&	$99\pm10$		& 	$200\pm21$		& $4.0\pm1.3$	&	This work \\
				&	89	&	8.0/0.13		&	$26.1\pm2.0$		&	$1.5\pm0.2$	&	$61\pm4$		&	$392\pm26$		& $4.7\pm0.7$	&	 \citet{ELR2021} \\
Circinus		&	53	&	5.0/0.09		&	$128.7\pm9.1$		&	$2.3\pm0.1$	&	$142\pm1$		&	$2960\pm107$	& $<0.4$	&	This work\\
				&	89	&	8.0/0.15		&	$225.4\pm9.1$		&	$1.8\pm0.1$	&	$151\pm1$		&	$4093\pm133$	& $<0.4$	&	This work \\
				&	214	&	18/0.34			&	$100.4\pm5.7$		&	$1.9\pm0.1$	&	$151\pm1$		&	$1908\pm74$		& $<0.4$	&	This work \\
Cygnus~A 		&	53	&	5.0/5.55		&	$2.2\pm0.2$			&	$11\pm3$	&	$133\pm8$		&	$242\pm21$		& $11\pm3$	&	\citet{ELR2018b}\\
				&	89	&	8.0/8.88		&	$2.1\pm0.2$			&	$9\pm2$		&	$129\pm7$		&	$189\pm17$		& $9\pm2$	&	\citet{ELR2018b}\\
Mrk~231 		&	89	&	8.0/6.94		&	$26.7\pm2.1$		&	$0.5\pm0.2$	&	$73\pm8$		&	$134\pm10$		& $0.5\pm0.2$	&	This work \\
NGC~1068 		&	53	&	5.0/0.35		&	$71.8\pm14.3$		&	$\sim1.2$	&	$\sim33$		&	$\sim862$		& $<0.4$	&	This work\\
				&	89	&	8.0/0.56		&	$87.7\pm14.3$		&	$\sim0.8$	&	$\sim141$		&	$\sim702$		& $<0.4$	&	This work\\
NGC~1275		&	89	&	8.0/2.86		&	$5.3\pm0.2$			&	$1.9\pm0.3$	&	$51\pm2$		&	$101\pm4$		& $1.9\pm0.3$	&	This work \\
NGC~4151		&	89	&	8.0/0.70		&	$3.4\pm0.4$			&	$<0.4$		&	-				&	-				& 	$<0.4$	&	This work \\
NGC~5506		&	89	&	8.0/1.02		&	$7.0\pm0.2$			&	$<0.4$		&	-				&	-				& $<0.4$	&	This work \\
PKS~1345+125 	&	89	&	8.0/18.19		&	$1.4\pm0.1$			&	$9.3\pm1.6$	&	$131\pm5$		&	$130\pm9$		& $9.3\pm1.6$	&	This work \\
\enddata
\tablenotetext{a}{For Centaurus A, $PA_{\rm int,E} = 45\pm9^{\circ}$ and $45\pm4^{\circ}$ at $53$ and $89$ \um, respectively.}
\end{deluxetable*}

We performed OTFMAP polarimetric observations in a sequence of four Lissajous scans, where each scan has a different halfwave plate (HWP) position angle (PA) in the following sequence: $5^{\circ}$, $50^{\circ}$, $27.5^{\circ}$, and $72.5^{\circ}$. In this new HAWC+ observing mode, the telescope is driven to follow a parametric curve with a nonrepeating period whose shape is characterized by the relative phases and frequency of the motion. Each scan is characterized by the scan amplitude, scan rate, scan angles, and scan duration.  A summary of the observations is shown in Table \ref{tab:table1}. We reduced the data using the Comprehensive Reduction Utility for SHARC II v.2.50-1 \citep[\textsc{crush};][]{kovacs2006,kovacs2008} and the \textsc{hawc\_drp\_v2.7.0} pipeline (i.e. \textsc{sofia\_redux\_v1.2}) developed by the data reduction pipeline group at the SOFIA Science Center. Each scan was reduced using \textsc{crush}, which estimates and removes the correlated atmospheric and instrumental signals, solves for the relative detector gains, and determines the noise weighting of the time streams in an iterated pipeline scheme. Each reduced scan produces two images associated with each array. Both images are orthogonal components of linear polarization at a given HWP PA.  We estimated the Stokes $IQU$ parameters using the double difference method in the same manner as the standard chop-nod observations carried out by HAWC+ described in Section 3.2 by \citet{Harper2018}. The degree ($P$) and $PA$ of polarization were corrected by instrumental polarization ($IP$) estimated using OTFMAP polarization observations of planets. To ensure the correction of the PA of polarization of the instrument with respect to the sky, we took the scans with a fixed line-of-sight (LOS) of the telescope. We rotated the Stokes $QU$ from the instrument to the sky coordinates. The polarization fraction was debiased and corrected by polarization efficiency. The final Stokes $IQU$, $P$, $PA$, polarized intensity ($PI$), and their associated errors were calculated and re-sampled to the half-beam size (Nyqvist sampling) at each wavelength.  The final images have a total on-source time as shown in Table \ref{tab:table1}. We estimate the observing time overhead to be 1.08. Several scans of Mrk~231, NGC~1275, and NGC~4151 were removed (Table \ref{tab:table1}) due to tracking error issues that occurred during the observations.

To complete the polarimetric AGN sample, we gather the polarimetric observations of already published objects using HAWC+ of Centaurus A \citep{ELR2021}, Circinus \citep{ELR2022a}, Cygnus A \citep{ELR2018b}, and NGC 1068 \citep{ELR2020,ELR2022a}. These observations were taken under programs 05\_0071 (PI: Lopez-Rodriguez, E.), 08\_0012 (PI: Lopez-Rodriguez, E. and Mao, S. A.), and Guaranteed Time Observations (GTO, PI: Dowell, C. D.), respectively. Table \ref{tab:table2} shows the complete sample used in this manuscript with their radio and MIR nuclear properties.

We performed aperture photometry of our sample within the central PSF at each wavelength. For point sources, we estimated the fluxes using an aperture (i.e., diameter) equal to $3.5\times$FWHM of the PSF (see Appendix \ref{app:CenA} for examples of PSF at $53$ and $89$ \um). For the extended sources, we used aperture correction within the beam size centered at the peak pixel of the AGN, except for Centaurus A. The aperture photometry was estimated using a known PSF from archival observations of planets. The known PSF was scaled such that the central beam has the same measured flux as the AGN, then the flux within the scaled-PSF using an aperture equal to $3.5\times$FWHM was estimated. The PSF of HAWC+ is known to be stable across observations \citep{Harper2018,ELR2022a}. Because the polarimetric observations require large integration times, the signal-to-noise ratio (SNR) of the total intensity was estimated to be $\ge500$. This SNR ensures a) low noise ($\le3$\% in total intensity) in an aperture equal to the beam of the observations and b) high accuracy ($\le1$ px) in the determination of the centroids of the object. For Centaurus A, we performed the PSF-fitting approach as described in Section \ref{subsec:IP} and Appendix \ref{app:CenA} to optimize the polarization measurement from the obscured PSF. Stokes $IQU$ and their uncertainties were estimated as follows. The values within the aperture for each Stokes $IQU$ were added. Then, the Stokes $IQU$ uncertainties associated with each pixel within the aperture were added in quadrature. In addition, the standard deviation in a region of the array without emission from the source was measured. HAWC+ provides flux calibration uncertainties $<10$\%, and PSF-variation was not found during the observations. In addition, the measured fluxes using HAWC+ observations are in good agreement, $<15$\%, with the fluxes obtained using \textit{Herschel} data (Figure \ref{fig:fig1}). The final uncertainties for each Stokes $IQU$ were computed as the sum in quadrature of these two quantities. The $P$, $PA$, and $PI$ were computed as described above. Table \ref{tab:table3} summarizes the measurements of the Stokes $I$, $P$, $PA_{\rm E}$ (E-vector), and $PI$ for our sample.


\section{Total and Polarized SEDs of AGN}\label{sec:Analysis}


\subsection{Total SEDs}\label{subsec:SED}

\begin{figure*}[ht!]
\includegraphics[scale=0.62]{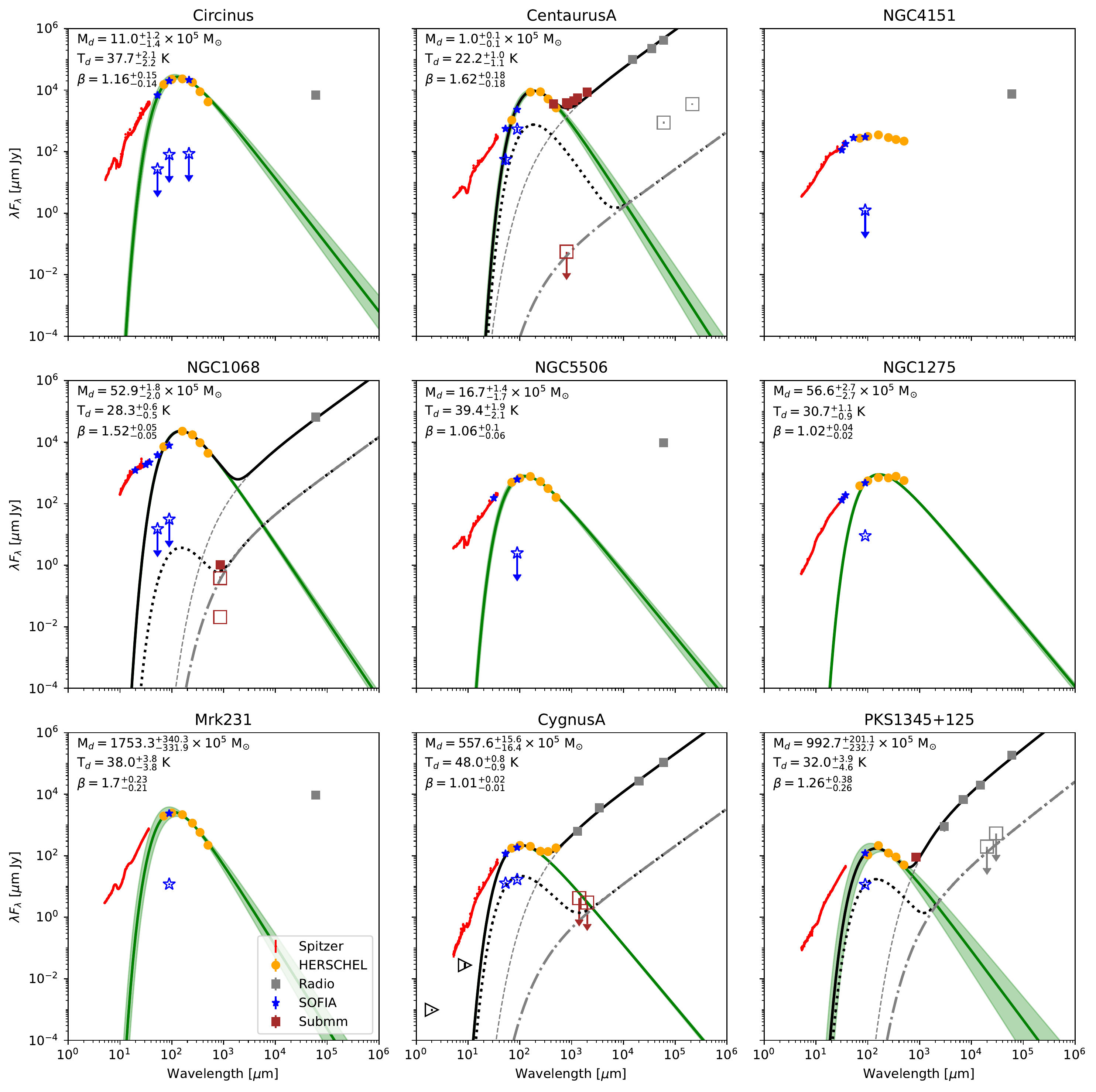}
\caption{$5$ \um\ - $5$ GHz total and polarized nuclear SED of AGN in order of redshift. Total SED (filled symbols) using \textit{Spitzer}, PACS/SPIRE/\textit{Herschel}, FORCAST/HAWC+/SOFIA and radio data as shown in the legend of the bottom left panel. 
Besides the differences in the radio emission from these sources, the $10-1000$ \um\ modified blackbody function (green line, Eq. \ref{eq:mBB}) shows a similar bump at $\lambda_{d}=100\pm20$ \um\ arising from dust emission with a characteristic dust temperature of $T_{\rm d}=34.8\pm7.8$ K (Section \ref{subsec:TIR}). The best-fit parameters of the modified blackbody functions are shown in each panel.
A fiducial synchrotron emission function (dashed grey line, Eq. \ref{eq:Fsyn}) with a slope of $\alpha_{\rm tk} = +0.21$ and $\alpha_{\rm th} = -0.9$ in the optically thick and thin regimes, respectively with a turnover wavelength at $543$ \um\ for Cygnus A and Centaurus A, and $2500$ \um\ for PKS~1345+125 (Section \ref{subsec:NTIR}) is shown. 
The total flux SED model (black solid line) is shown. 
Polarized flux SED (open symbols) using NIR (black triangle), SOFIA (open star), and sub-mm (open squares) are shown. 
The scaled synchrotron (green dashed-dotted line) and thermal (black dotted line) polarized SEDs are shown for the polarized AGN (Section \ref{subsec:pSED}). The IR polarized SED of Centaurus A, Cygnus A, and PKS~1345+125 follow the shape of the IR bump in the total flux SED with a negligible contribution of synchrotron emission.
 \label{fig:fig1}}
\end{figure*}

To support the analysis of the origin of the polarized thermal emission in our AGN sample, we compile the total nuclear SEDs from $5$ \um\ to $5$ GHz. The SEDs have comparable, or better, angular resolution to that from our HAWC+ observations (Table \ref{tab:table3}). The angular resolution is high for all observations to separate the AGN from extended emission and/or radio lobes associated with the host galaxies. In the $10-30$ \um\ wavelength range, the \textit{Spitzer} spectroscopic data from the Combined Atlas of Sources with \textit{Spitzer} IRS Spectra (CASSIS) archive\footnote{CASSIS spectra can be found at \url{https://cassis.sirtf.com/atlas/welcome.shtml}} were used. We took the low-resolution and optimal extraction spectra \citep{Lebouteiller2011}. In the $70-500$ \um\ wavelength range, we took PACS and SPIRE/\textit{Herschel} observations from the \textit{Herchel} archive\footnote{\textit{Herschel} Archive can be found at \url{http://archives.esac.esa.int/hsa/whsa/.}}. The FORCAST and HAWC+ observations from SOFIA were used to complement the $30-214$ \um\ wavelength range. The total flux SOFIA observations of these datasets are described in \citet[][Fuller et al. in prep.]{Fuller2016,Fuller2019}. New total and polarimetric data associated with HAWC+ are described in Section \ref{sec:OBS}. In the $30-500$ \um\ wavelength range, we performed aperture photometry of the SOFIA and \textit{Herschel} observations using a diameter equal to the beam size at each wavelength as described in Section \ref{sec:OBS}. References for the nuclear fluxes at $5$ GHz are shown in Table \ref{tab:table2}.

\begin{deluxetable*}{lccccccccc}[ht!]
\tablecaption{Best fit parameters of the modified blackbody function (Eq. \ref{eq:mBB}).
(a) Object name. 
(b) Dust mass in units of $\times 10^{5}$ M$_{\odot}$. 
(c) Characteristic dust temperature in units of K. 
(d) Dust emissivity index. 
(e) Total infrared luminosity in units of erg s$^{-1}$.
(f) $89$ \um~luminosity in units of erg s$^{-1}$.
(g) The radius of the optically thick dusty environment assuming radiative equilibrium in units of pc using Eq. \ref{eq:rd}. 
(h) The radius of the optically thin dusty environment assuming radiative equilibrium in units of pc using Eq. \ref{eq:rd_th}.
(i) Spatial scale of the beam size of the observations at $89$ \um~(Table \ref{tab:table3}).
\label{tab:table4}}
\tablecolumns{7}
\tablewidth{0pt}
\tablehead{\colhead{Object}& \colhead{M$_{\rm d}$} 	&	\colhead{T$_{\rm d}$}		&	 \colhead{$\beta$}	& \colhead{$\log_{10}(L_{\rm TIR})^{\star}$} & \colhead{$\log_{10}(L_{\rm 89\mu m})$} & \colhead{$r_{\rm d, tk}$} & \colhead{$r_{\rm d, th}$} & \colhead{$\theta_{\rm{89 \mu m}}$} \\
		&	\colhead{($\times10^{5}$ M$_{\odot}$)}  &	\colhead{(K)} & 	&	\colhead{(erg s$^{-1}$)} &	\colhead{(erg s$^{-1}$)} 	&	\colhead{(pc)} &	\colhead{(pc)} &	\colhead{(pc)} \\
\colhead{(a)} & \colhead{(b)} & \colhead{(c)} & \colhead{(d)} & \colhead{(e)} & \colhead{(f)} & \colhead{(g)} & \colhead{(h)}  & \colhead{(i)} 
}
\startdata
Centaurus A		& $1.0^{+0.1}_{-0.1}$			& $22.2^{+1.0}_{-1.1}$	& $1.62^{+0.18}_{-0.18}$	& $43.38^{+0.06}_{-0.05}$ 	& $41.09^{+0.14}_{-0.12}$	& $4.7^{+0.4}_{-0.4}$		& $457^{+217}_{-167}$		& $130$	\\
Circinus			& $11.0^{+1.2}_{-1.4}$			& $37.7^{+2.1}_{-2.2}$	& $1.16^{+0.15}_{-0.14}$	& $44.22^{+0.10}_{-0.09}$ 	& $42.15^{+0.11}_{-0.11}$ 	& $5.3^{+0.6}_{-0.5}$		& $185^{+43}_{-70}$		& $150$ \\
Cygnus~A 		& $557.6^{+15.6}_{-16.4}$		& $47.7^{+0.9}_{-0.9}$	& $1.01^{+0.02}_{-0.01}$	& $45.79^{+0.03}_{-0.03}$ 	& $43.69^{+0.03}_{-0.03}$ 	& $20.3^{+0.4}_{-1.8}$		& $417^{+36}_{-37}$			& $8880$ \\
Mrk~231 			& $1753.3^{+340.3}_{-331.9}$		& $38.0^{+3.8}_{-3.8}$	& $1.70^{+0.23}_{-0.21}$	& $46.57^{+0.18}_{-0.16}$ 	& $44.53^{+0.17}_{-0.17}$	& $78.6^{+19.7}_{-13.4}$		& $5802^{+6773}_{-1699}$	& $6940$ \\
NGC~1068 		& $52.9^{+1.8}_{-2.0}$			& $28.3^{+0.6}_{-0.5}$	& $1.52^{+0.05}_{-0.05}$	& $45.07^{+0.03}_{-0.03}$	& $42.96^{+0.05}_{-0.04}$	& $23.8^{+1.8}_{-0.8}$		& $1736^{+213}_{-158}$		& $560$ \\
NGC~1275		& $56.6^{+2.7}_{-2.7}$			& $30.7^{+1.1}_{-0.9}$	& $1.02^{+0.04}_{-0.02}$	& $44.99^{+0.02}_{-0.03}$	& $42.84^{+0.04}_{-0.04}$	& $18.3^{+0.9}_{-0.8}$		& $440^{+62}_{-7}$	 		& $2860$ \\
NGC~4151$^{\dagger}$	& - & - & - & - & - & - & - & -\\
NGC~5506		& $16.7^{+1.4}_{-1.7}$			& $39.4^{+1.9}_{-2.1}$	& $1.06^{+0.10}_{-0.06}$	& $44.37^{+0.05}_{-0.07}$ 	& $42.29^{+0.06}_{-0.07}$	& $5.8^{+0.6}_{-0.6}$		& $146^{+19}_{-35}$	 		& $1020$ \\
PKS~1345+125 	& $992.7^{+201.1}_{-232.7}$		& $32.0^{+3.9}_{-4.6}$	& $1.26^{+0.38}_{-0.26}$	& $46.29^{+0.16}_{-0.21}$ 	& $44.19^{+0.21}_{-0.32}$	& $63.4^{+47.4}_{-12.8}$		& $3346^{+3967}_{-1784}$	& $18190$ \\
\enddata
\tablenotetext{^\dagger}{A single modified blackbody within the $30-500$ \um\ wavelength range for NGC~4151 does not reproduce the total flux SED.}
\tablenotetext{^\star}{Thermal infrared luminosity in the $8-1000$ \um\ wavelength range.}
\end{deluxetable*}

Figure \ref{fig:fig1} shows the compiled $5$ \um\ $-$ $5$ GHz nuclear total SEDs of the AGN sample. All targets in the sample have similar $10-1000$ \um\ SED behavior. All total SEDs show a bump at $\sim100$ \um\ arising from thermal dust emission from the core (Section \ref{subsec:TIR} and \ref{subsec:NTIR}). This IR bump has been identified as arising from dust emission with a steep decrease towards sub-mm and near-IR wavelengths \citep[i.e.,][]{Barvainis1992}. The main difference between the continuum total SEDs resides in the radio emission from their jets. This result is in agreement with the classical definition of RL and RQ AGN and their total SEDs \citep[e.g.,][]{Antonucci1993,Elvis1994,UP1995}. Thus, the $1-1000$ \um\ total SEDs range does not necessarily provide further information about the nature of RL and RQ AGN. We find that the opposite is true in the FIR polarized flux (Section \ref{subsec:pSED}).

\subsubsection{Thermal emission component}\label{subsec:TIR}

We characterize the observed IR bump of the total SEDs shown in Figure \ref{fig:fig1}. The dust properties from $20-500$ \um\ are estimated using a single-temperature modified blackbody function expressed as

\begin{equation}\label{eq:mBB}
F_{\lambda} = \frac{M_{\rm d}}{D_{\rm L}^{2}}k_{\lambda_{0}}\left(\frac{\lambda}{\lambda_{0}}\right)^{-\beta}B_{\lambda}(T_{\rm d})
\end{equation}
\noindent
where M$_{\rm d}$ is the dust mass, $D_{\rm L}$ is the luminosity distance to the source, $k_{\lambda_{0}}$ is the dust mass absorption coefficient to be $0.29$ m$^{2}$ kg$^{-1}$ at a wavelength of $\lambda_{0} = 250$ \um\ \citep{Wiebe2009}, $\beta$ is the dust emissive index, and $B_{\lambda}(T_{\rm d})$ is the blackbody function at a characteristic dust temperature of $T_{\rm d}$.

We have three free model parameters: $M_{\rm d}$, $\beta$, and $T_{\rm d}$ that we fit within the $20-500$ \um\ wavelength range. We compute a   Markov Chain Monte Carlo (MCMC) approach using the differential evolution metropolis sampling step in the \textsc{python} code \textsc{pymc3} \citep{pymc}. The prior distributions are set to flat within the range of $M_{\rm d} = [10^{3}-10^{11}]$ M$_{\odot}$, $\beta = [1,2]$, and $T_{\rm d} = [10,100]$ K. We run the code using 5 chains with 5,000 steps and a 1,000 burn-in per chain, which provides 25,000 steps for the full MCMC code useful for data analysis. Final median values and $1\sigma$ uncertainties are shown in Table \ref{tab:table4}, and the best-fit model with the associated $1\sigma$ uncertainty is plotted for each source in Figure \ref{fig:fig1}. 

We find that the IR bump peaks within the wavelength range of $[77,142]$ \um, with a median of $\lambda_{\rm d} = 100\pm20$ \um. The characteristic dust temperature of the IR bump is in the range of $[22.2, 47.7]$ K, with a median of $T_{\rm d} = 34.8\pm7.8$ K. The dust emissive index is in the $\beta = [1.01-1.70]$ range, with a median of $\beta = 1.21\pm0.26$. This result is in agreement with the typically assumed value of $1.5$ for galaxies in the range of $1-2$ \citep{H1983}. The associated dust mass of the IR bump has a wide range of $1.0\times10^{5}-1.8\times 10^{8}$ M$_{\odot}$. This is mainly due to the volume associated with the measured core, given the fixed angular resolution of our observations. The beam sizes at $89$ \um~vary from $130$ pc in Centaurus A to $18.19$ kpc in PKS~1345+125. Also, the differences of AGN types, for example, Mrk~231 is an ultraluminous infrared galaxy with known dust masses up to $\sim10^{9}$M$_{\odot}$ \citep{Lisenfeld2000}. We estimate the thermal IR luminosity, $L_{\rm TIR}$, in the range of $8-1000$ \um\ to be in the range of $\log_{10} (L_{\rm TIR}~\mbox{[erg s$^{-1}$]}) = [43.38-46.57]$.  

We compute the effective size of the dust emitting at $89$ \um~assuming radiative equilibrium at a distance, $r_{\rm d}$, from the central engine as

\begin{equation}\label{eq:rd}
r_{\rm d, tk} = 441\left( \frac{30~\rm{K}}{T_{\rm d}}\right)^{2} \left( \frac{L_{89 \mu m}}{10^{12}L_{\odot}}\right)^{1/2} ~\mbox{pc}
\end{equation}
\noindent
\citep{Scoville2013} for a normalized dust temperature of $30$ K, and a luminosity at $89$ \um\ normalized at $10^{12}L_{\odot}$ (Table \ref{tab:table4}). Note that this relation assumes that the emission arises from optically thick dust. Our best fits suggest that a modified blackbody function best describes the IR bump, which indicates that the dust is optically thin at FIR wavelengths (Section \ref{subsec:ThermalPol}). The effective source size from optical thin dust can be estimated as

\begin{equation}\label{eq:rd_th}
r_{\rm{d,th}} = \left(\frac{1300~\rm{K}}{T_{\rm d}}\right)^{\frac{4+\beta}{2}} \left( \frac{L_{89 \mu m}}{10^{46} \rm{erg~s^{-1}}} \right)^{1/2}
\end{equation}
\noindent
\citep[i.e.,][]{Netzer2008,BN2019}. We show the radii arising from a structure with optically thick and optically thin dust in Table \ref{tab:table4}. For optically thick dust, $r_{\rm{d,tk}}$ is in the range of $[5,79]$ pc, and $[146, 5802]$ pc for optically thin dust, $r_{\rm{d,th}}$. We discuss these results in Section \ref{subsec:Source}.

A single-temperature modified blackbody source can explain the IR bump of all galaxies, except for NGC~4151, whose IR total SED is flatter than the other galaxies in our sample. NGC~4151 is a type 1.5 AGN with an extended ($\sim$ 3.5\arcsec, $\sim$255.5 pc) dusty structure cospatial with the narrow-line region. This structure was observed in total intensity at 10 \um\ (broad-band filter N) with Gemini-North/OSCIR \citep{Radomski2003}. Follow-up imaging- and spectro-polarimetric observations with narrow-band filters at 8.7, 10.3, and 11.6 \um\ using CanariCam on the 10.4-m Gran Telescopio CANARIAS (GTC) measured a similar structure in total flux. This extended structure is polarized at a level of $0.5-1$\% potentially arising from dust scattering \citep{ELR2018c}. Given the complexity of the extended emission within our beam size ($8$\arcsec, $580$ pc), several blackbodies may be required to explain the IR total SED of NGC~4151. As we are interested in the AGN, a detailed model of NGC4151 is outside the scope of our goals.

\subsubsection{Non-thermal emission component}\label{subsec:NTIR}

We follow a similar approach from \citet{ELR2018b} to characterize the non-thermal emission from the total SED of AGN. We assume a self-absorbed synchrotron component with a flat spectrum, $\nu^{\alpha_{\rm tk}}$, in the optically thick region and a turnover at a frequency, $\nu_{\rm c}$, with a flat spectrum, $\nu^{\alpha_{\rm th}}$, in the optically thin region. The synchrotron emission is described as

\begin{equation}\label{eq:Fsyn}
F_{\rm syn} = C \left(\frac{\nu}{\nu_{\rm c}}\right)^{\alpha_{\rm tk}} e^{\left( -\frac{\nu}{\nu_{\rm c}} \right)^{\alpha_{\rm th}}}
\end{equation}
\noindent
where $C$ is a scale factor. 

Our goal is to show the overall tendency of the synchrotron component rather than a fit to the SEDs. The total flux $F_{\rm syn}$ is scaled so that it is equal to the total flux at $5$ GHz, and the polarized flux $PI_{\rm syn}$ is equal to the upper limits at radio wavelengths (Section \ref{subsec:pSED}). We adopt the same spectrum $\alpha_{\rm tk} = +0.21$ and $\alpha_{\rm th} = -0.9$ from the best fit of Cygnus A \citep{ELR2018b}. The turnover frequency is adopted to be $543$ \um\ from the best fit of Cygnus A \citep{ELR2018b} and Centaurus A, and $2500$ \um\ for  PKS~1345+125. These turnover frequencies ensure that the synchrotron emission is always equal to or smaller than the total flux at $500$ \um, whose emission may arise from a contribution of both thermal and non-thermal components. For Centaurus A, the break frequency is required to satisfy the upper limit in polarized SED (brown open square). A straight line will produce a higher polarized flux at $850$ \um~than the measurement. Note that even a single power law towards short wavelengths would produce negligible synchrotron emission at $100$ \um.

The total SED model is computed as the linear combination of the modified blackbody function (Eq. \ref{eq:mBB}) and the synchrotron component (Eq. \ref{eq:Fsyn}), i.e. $F_{\rm T} = F_{\lambda} + F_{\rm syn}$. As shown in Figure \ref{fig:fig1}, the synchrotron emission is negligible at $\sim100$ \um.

\subsection{Intrinsic FIR polarization}\label{subsec:IP}

Our FIR polarimetric measurements may be diluted by the emission and/or absorption from the dust lane in the host galaxy. For extended sources, we quantify the polarized emission arising from the host galaxy and subtract it from the measured polarization to compute the intrinsic polarization of the nuclear source. Here, we show the computation for Centaurus A, Circinus, and NGC 1068. The rest of the galaxies are point-like sources and the polarized emission from aligned dust grains in the host galaxy cannot be estimated. However, the integrated FIR polarization contribution from the host galaxy is considered to be negligible (Section \ref{subsec:ThermalPol}). Thus, we take their measured polarization, $P_{\rm m}$, shown in Table \ref{tab:table3}, as the intrinsic polarization.

\textbf{Centaurus A.} \citet{ELR2021} found a $89$ \um\ polarized nuclear source with a measured polarization of $P_{\rm m} = 1.5\pm0.2$\% at a $PA_{\rm E} = 61\pm4^{\circ}$ ($PA_{\rm B}=151\pm4^{\circ}$). The $PA_{\rm B}$ is the B-field orientation after a rotation of $90^{\circ}$ of the E-vector, i.e. $PA_{\rm B} = PA_{\rm E} + 90^{\circ}$ (Section \ref{subsec:ThermalPol}). The AGN is embedded in a warped molecular disk with a measured polarization of $P_{\rm gal}\sim3.5$\% and a PA in the range of $PA_{\rm gal,E} = [15-57]^{\circ}$ ($PA_{\rm gal,B}=[105-147]^{\circ}$). Thus, the measured nuclear polarized emission may be the contribution from the AGN and the host galaxy. 

The nuclear total flux can be decomposed by emission from the AGN and host galaxy. To obtain the fractional contribution to the total emission from both unresolved (AGN) and extended (host galaxy) components, we follow a similar procedure applied to the HAWC+/SOFIA observations of NGC1068 by \citet{ELR2018a}. First, the fluxes in a circular aperture equal to the FWHM at $53$ and $89$ \um\ were measured. This flux represents the total flux, $F_{\rm T}$, of the nuclear resolved source at the given wavelength and minimizes the contribution from the extended diffuse emission (Table \ref{tab:table3}). Second, the central $3\times$FWHM radius emission was fitted with a composite model using the corresponding HAWC+ point-spread-function (PSF) to each observation and a 2D Gaussian profile. The total flux from the scaled-PSF, $F_{\rm PSF}^{\rm m}$, represents the maximum likely contribution from an unresolved nuclear component at the given angular resolution of the observations. The total flux from the 2D Gaussian profile, $F_{\rm gal}^{\rm m}$, provides the minimum contribution of the extended component surrounding the central source. The total flux model, $F_{\rm T}^{\rm m}$, is the sum of both  PSF and 2D Gaussian profiles. This method has four free parameters: the amplitude of the PSF, $C_{\rm PSF}$, the FWHM of the long, $b$, and short, $a$, axes and the PA, $\theta$, of the 2D Gaussian profile. The amplitude of the 2D Gaussian profile is estimated as the peak of the observed total intensity at each wavelength minus the amplitude of the PSF. This approach ensures that the final model has the same peak flux as the observations.

The fitting routine was performed using a Markov Chain Monte Carlo approach using the differential evolution metropolis sampling step in the \textsc{python} code \textsc{PyMC3} \citep{pymc}. The prior distributions are set to flat within the ranges of $C_{\rm PSF} =[0,\max({Flux})]$, where the upper limit corresponds to the peak flux at $53$ and $89$ \um; FWHM$_{\rm a,b} = [2,10]\times$FWHM$_{53,89}$, where FWHM$_{53,89}$ corresponds to the FWHM at $53$ and $89$ \um; and $\theta = [0,180)^{\circ}$. Note that the 2D Gaussian profiles are set to be always larger than the PSF at a given wavelength. We run the code using 10 chains with 5,000 steps and a 1,000 burn-in per chain, which provides 50,000 steps for the full MCMC code useful for data analysis. For the best model, the nuclear fluxes of the PSF, 2D Gaussian, and final model within the beam size at each wavelength were estimated. Appendix \ref{app:CenA} shows the best-inferred parameters and associated fluxes (Table \ref{tabApp:table1}), the posterior distributions of the free parameters (Fig. \ref{figApp:fig1}), and the original and residual images of the PSF-fitting routine (Fig. \ref{figApp:fig2}). 

Our PSF-fitting approach shows that the AGN contributes at $47\pm5$ \% and $32\pm3$ \% within the central $80$ and $130$ pc at $53$ and $89$ \um, respectively. We compute the polarization of the nuclear source corrected by host galaxy dilution as

\begin{equation}
P_{\rm agn} = P_{\rm m} \left( 1+ \frac{F_{\rm gal}}{F_{\rm PSF}} \right)
\label{eq:Pflux}
\end{equation}
\noindent
\citep[][]{ELR2013,ELR2015}, where $P_{\rm m}$ is the measured polarization, $F_{\rm gal}$ is the extended flux of the host galaxy, and $F_{\rm PSF}$ is the flux of the unresolved core. 

After $P_{\rm agn}$ is estimated, this value and the PA$_{\rm obs}$ are corrected by the vector quantity arising from the host galaxy as follows. The Stokes parameters $q_{\rm gal}$ and $u_{\rm gal}$ from the host galaxy (P$_{\rm gal}$ and PA$_{\rm gal}$) are estimated. Then, they are subtracted from the Stokes parameters $q_{\rm agn}$ and $u_{\rm agn}$, which are estimated using the corrected polarization as shown in Eq. \ref{eq:Pflux} and the observed PA as shown in Table \ref{tab:table3}. Thus, the final intrinsic Stokes parameters are estimated such as $q_{\rm int} = q_{\rm agn} - q_{\rm gal}$ and $u_{\rm int} = u_{\rm agn} - u_{\rm gal}$. We estimate an intrinsic polarization of $P_{\rm int} = 4.0\pm1.3$\% and $4.7\pm0.7$\% with a PA$_{\rm int,E} = 45\pm9^{\circ}$ and $45\pm4^{\circ}$ at $53$ and $89$ \um, respectively. The B-field orientation is estimated to be PA$_{\rm int,B} = 135\pm9^{\circ}$ and $135\pm4^{\circ}$ at $53$ and $89$ \um, respectively.

\textbf{Circinus.} Within the $53-214$ \um\ wavelength range, the measured nuclear polarization is estimated to be $\sim2$\% with a PA$_{\rm E}\sim135^{\circ}$ \citep{ELR2022a}, where the B-field orientation, PA$_{\rm B}\sim45^{\circ}$, is similar to the orientation of the central $5$ kpc bar at a PA of  $\sim45^{\circ}$ \citep{Jones1999}. Based on the B-field model by Grosset et al. (in prep.), the polarization arises from the spiral arms and starburst ring within the central $1$ kpc. These authors found that the AGN at each wavelength is consistent with an unpolarized source. We take that the AGN is intrinsically unpolarized at FIR wavelengths and take an upper limit of $0.4$\% equal to the absolute minimum polarization measured by HAWC+ \citep{Harper2018}.

\textbf{NGC 1068.}  Although the 89 \um\ polarization along the $1$ kpc bar is measured to be in the range of $1-3$\% \citep{ELR2018b}, a minimum in the polarization is found at the nucleus with a PA$_{\rm E}\sim141^{\circ}$. The inferred B-field orientation, PA$_{B}\sim39^{\circ}$, is similar to the orientation of the bar, PA $\sim48^{\circ}$ \citep{Scoville1988}. At 53 \um, we measure a nuclear polarization $\sim1.2$\% at a PA$_{\rm E}\sim33^{\circ}$ within the central beam ($5$\arcsec, $70$ pc). Note that the $PA_{\rm E}$ of polarization rotates by $\sim119^{\circ}$ between $53$ and $89$ \um. This result can be easily observed in figure 1 by \citet{ELR2022a}. This angular rotation may suggest that the thermal polarization at $53$ \um\ arises from absorptive polarization, which indicates that the measured PA of polarization is the orientation of the B-field in the bar. Indeed the PA of polarization of $33^{\circ}$ arising from absorptive polarization at $53$ \um\ is in agreement with the inferred B-field of $39^{\circ}$ arising from emissive polarization at $89$ \um. Another plausible explanation is that the measured PA of polarization at $53$ \um\ may arise from polarized thermal emission at the origin of the spiral arms coming out from the core. The angular resolution improvement from $560$ to $350$ pc at 89 and 53 \um\, respectively, may be able to explain the angular difference. The $8-12$ \um~imaging polarimetric observations using Canaricam on the 10.4-m GTC within the central $\sim$100 pc of NGC~1068 showed a highly polarized, $\sim4-7$\%, region  with a $PA _{\rm E}\sim44^{\circ}$ at $\sim24$ pc north from the core \citep{ELR2016}. The polarization was found to arise from dust and gas emission in the ionization cone, heated by the active nucleus and jet, and further extinguished by aligned dust grains in the host galaxy. In addition, these observations found an unpolarized, $\sim0.4$\%, core. Using ALMA  860 \um\ polarimetric observations at a resolution of $0\farcs07$ ($4.2$ pc), \citet{ELR2020b} measured an unpolarized central core and a polarized dusty torus at a level of $3.7\pm0.5$\% with a B-field orientation of $108\pm2^{\circ}$ within $5-9$ pc east from the core. Given the low angular resolution ($560$ pc) of the $89$ \um\ HAWC+ observations, we suggest that the nuclear FIR polarization arises from the polarized thermal emission of the aligned dust grains in the bar--rather than the emission from the AGN. We conclude that the AGN of NGC 1068 at $89$ \um\ is intrinsically unpolarized at FIR wavelengths. We take an upper limit of $0.4$\% as the absolute minimum polarization measured by HAWC+ \citep{Harper2018}. The $53$ \um\ polarimetric observations are inconclusive and will not be used in this work.

In summary, we find that at $89$ \um\ Circinus, Mrk~231, NGC~1068, NGC~4151, and NGC~5506 are consistent with unpolarized sources. We find that Centaurus~A, Cygnus~A, PKS~1345+125 (4C+12.50), and potentially NGC~1275 (Section \ref{subsec:DeltaPA}) are consistent with intrinsic polarized sources at a level of $1.9-11$\%. Thus, the $89$ \um~polarization fraction shows differences between AGN.

\subsection{Polarized SEDs}\label{subsec:pSED}

We compile the nuclear polarized SEDs from $2$ \um\ to $5$ GHz to support the analysis of the origin of the polarized thermal emission from AGN. The polarized SEDs have similar, or better, angular resolution to that from our HAWC+ observations. Appendix \ref{app:PolSED} (Table \ref{tab:table7}) shows the wavelengths, apertures, and polarization fluxes of the polarized measurements shown in Figure \ref{fig:fig1}. Due to the small number of data points in the polarized SEDs, we aim to provide fiducial polarized SEDs to quantify the relative contribution of thermal and non-thermal polarized emission at FIR wavelengths.

For the polarized AGN, we scaled the polarized thermal emission to be equal to the median polarized flux within the $53-89$ \um\ wavelength range. This scale factor is equal to the median polarization fraction of $4.5\%$ and $10$\% within $53-89$ \um\ i.e., $0.045$ for Centaurus A, and $0.1$ for Cygnus A and PKS~1345+125. We scaled the polarized non-thermal emission to be equal to the polarized fluxes at mm wavelengths. These scale factors are $0.004$, $0.001$, and $0.002$ for Centaurus~A, Cygnus~A, and PKS1345+125, respectively. These AGN have an intrinsic polarization fraction $<0.4$\% at mm wavelengths arising from synchrotron emission. Figure \ref{fig:fig1} shows the total polarized SEDs (black dashed line) for the polarized AGN. 

For the unpolarized AGN, we only show the polarized SED of NGC 1068. The reason is that this AGN is the only object that has the dusty torus and surrounded jet resolved within the central $20$ pc at a resolution of $4.2$ pc ($0\farcs07$) at $860$ \um\ \citep{ELR2021a}. These observations showed an unresolved core of  $4.2$ pc in diameter (i.e., FWHM) with an upper-limit polarization fraction of $<0.2$\%. Due to the optically thick dusty torus, the central unpolarized core arises from a self-absorbed synchrotron emission. The dusty torus is resolved and polarized at $3.7\pm0.5$\% at $860$ \um\ across the $5-9$ pc eastern region of the equational axis. Figure \ref{fig:fig3} shows the total flux of $12\pm0.5$ mJy (brown-filled square), the polarized flux of the dusty torus of $0.44\pm0.06$ mJy, and the unpolarized core with an upper-limit polarized flux of $<0.024$ mJy. We use the same total SED for the polarized AGN with a turnover wavelength of $2000$ \um. We scaled the polarized flux as follows. The non-thermal emission is equal to the upper-limit polarized flux of the unpolarized core, and the thermal emission is equal to the polarized dusty torus. The scale factors for thermal and non-thermal components are $0.0002$ and $0.008$, respectively. This result indicates that thermal and non-thermal components are unpolarized at $<0.02$\%, respectively, at $100$ \um. 

For the polarized AGN (i.e., Centaurus~A, Cygnus~A, and PKS1345+125) and unpolarized AGN (NGC~1068), we find that the polarized synchrotron emission is negligible at FIR wavelengths. Figure \ref{fig:fig1} shows that the non-thermal polarized emission (i.e., synchrotron component) is several orders of magnitude lower than the polarized thermal emission arising from magnetically aligned dust grains at FIR wavelengths.  This result holds even if a single power law is used to interpolate the synchrotron emission to FIR wavelengths. As we will discuss in Section \ref{subsec:DeltaPA}, NGC~1275 is excluded from this analysis because the origin of the measured polarization is inconclusive. For the polarized AGN with several data points in the FIR (i.e., Centaurus~A and Cygnus~A), the polarized IR emission has the same shape as the IR bump of the total SED. The polarized thermal component contributes $<0.4$\% of the total SEDs in our unpolarized AGN (i.e., Circinus, Mrk~231, NGC~1068, NGC~5506). 

We conclude that polarized thermal emission by means of magnetically aligned dust grains is the dominant polarization mechanism in the $53-214$ \um~wavelength range. By combining the total flux and polarized SEDs with the characterization of the IR bump, we can use the FIR polarization measurements to infer the B-field orientation of the dusty environment surrounding polarized AGN. We show the first observational evidence of an intrinsic difference between AGN using the $89$ \um~nuclear polarized SEDs.


\section{Radio-loudness}\label{sec:R}

\begin{figure*}[ht!]
\includegraphics[scale=0.65]{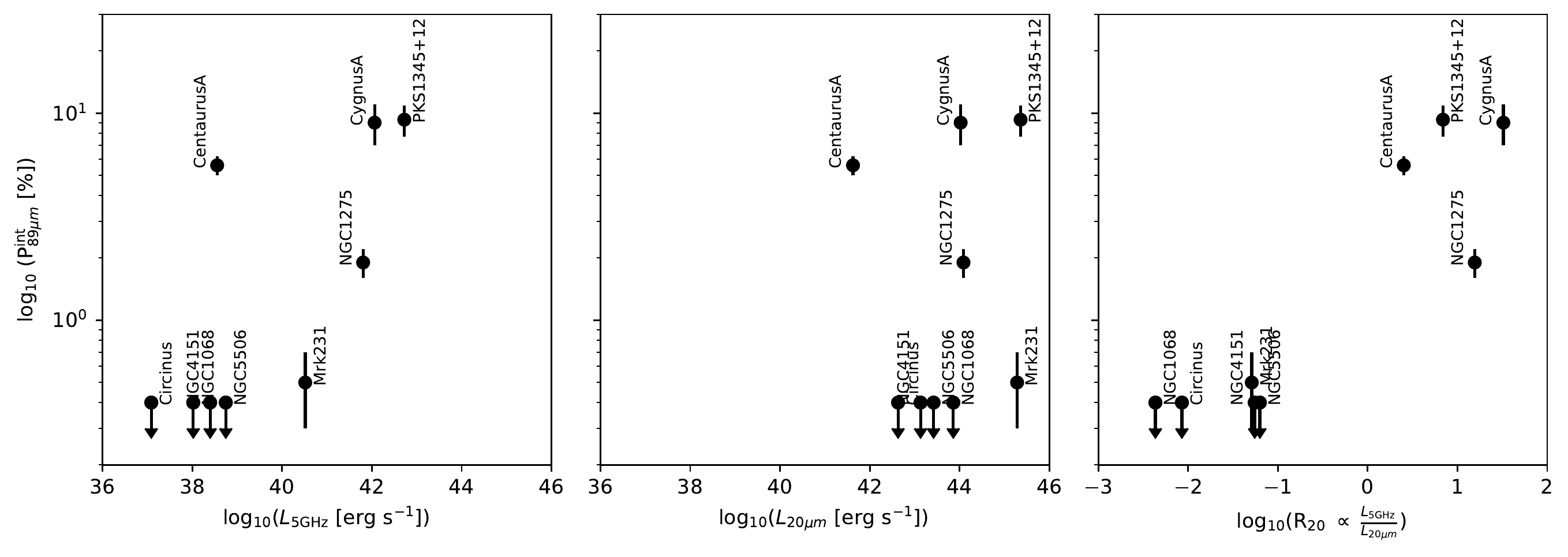}
\caption{$89$ \um\ intrinsic polarization vs. $5$ GHz (left) and $20$ \um\ (middle) nuclear luminosities, and radio-loudness, $R_{20}$ (right). We find that the thermal polarization by means of magnetically aligned dust grains increases with the nuclear radio-loudness, $R_{20}$.
 \label{fig:fig2}}
\end{figure*}

We estimate the radio-loudness of the objects in our sample to characterize the physical origin of the nuclear polarization between AGN. We look at the ratio of the radio fluxes, $F_{\rm 5GHz}$, to some proxy for the optical/ultraviolet (UV) continuum that works for both Type 1 and Type 2 AGN. Here, we would like to define this carefully according to the following precepts. We want to refer as much as possible only to the intrinsic properties of the AGN. We avoid quantities that may be likely to be beamed, such as self-absorbed cores of radio-loud objects. Those cores are the bases of jets, which are likely to be beamed. Our sample comprises galaxies whose compact cores rarely dominate and their radio images often have linear morphologies and are non-relativistic. We decided not to use the classical definition of radio loudness, $R=L_{\rm 5GHz}/ L_{\rm B}$. The reason is that $R$ is likely to suffer substantial extinction from circumnuclear dust. This is particularly important in the Type 2 objects where the broad-line region suffers from dust extinction.  $R$ may be an unreliable measure of radio-loudness.

We estimate the radio-loudness as $R_{20}=L_{\rm 5GHz}/L_{20\mu m}$, where $L_{20\mu m}$ is the nuclear luminosity at $20$ \um.  At MIR ($\sim8-13$ \um) wavelengths, except for the weakest AGN, the sub-arcsecond flux is dominated by a warm and nearly isotropically radiating point-like source heated by the nucleus \citep[i.e.][]{Asmus2014}. At wavelengths longer than $20$ \um, the thermal emission has been found to predominantly arise from the dusty torus \citep[i.e.][]{Hoenig2011,Fuller2016,ELR2018a}. Potentially, $L_{20}$ may suffer from contamination by the circumnuclear starburst in some objects (e.g., the ULIRGs Mrk231 and PKS1345+12). However, if starburst dominates, the dust polarized emission would be low (Section \ref{subsec:ThermalPol}). In addition, the $20$ \um\ fluxes precludes significant anisotropy given the constant ratios to the optical/UV in Type 1 AGN and to the X-rays in both AGN types \citep[i.e.,][]{Hoenig2011}. In principle, modest noise is introduced by a range of covering factors, perhaps a factor of two \citep{Lawrence2013}. Using the $20$ \um\ fluxes as a proxy of the core luminosity has several advantages. The $20$ \um\ emission is a) insensitive to reddening compared to the optical fluxes, b) averaged over the many sightlines through the dusty torus, and c) always measuring the radiation from the same wide-equatorial directions. These conditions imply that the IR emission can be used as a calorimeter \citep{A2012}.

The $20$ \um\ fluxes, F$_{20\mu m}$, were taken from the Sub-ArcSecond Mid-InfRared Atlas of Local AGN (SASMIRALA\footnote{The Sub-ArcSecond Mid-InfRared Atlas of Local AGN (SASMIRALA) can be found at \url{https://dc.zah.uni-heidelberg.de/sasmirala/q/cone/form}}). Specifically, we selected the `PSF'-fluxes obtained using a Gaussian fitting of the central beam of the observations, typically $\sim0.4$\arcsec\ using $10$-m class telescopes \citep{Asmus2014}. The $11.6$ \um\ flux of Mrk~231 was taken from \citet{ELR2017} using the Gran Telescopio CANARIAS, and the $20$ \um\ flux of PKS~1345+125 was taken from \citet{Veilleux2009} using \textit{Spitzer}. These two galaxies are point-like sources, and the aperture photometry does not require Gaussian fitting procedures. The $11.6$ \um\ fluxes were multiplied by a factor of $2$ to account for the typical change of fluxes in Type 2 AGN \citep[i.e.][]{Hoenig2011,Asmus2014}. 

The luminosity at $20$ \um\ is estimated as

\begin{equation}\label{eq:L20}
L_{20\mu m} = 4 \pi D_{\rm L}^{2} \nu_{20\mu m} F_{20\mu m}
\end{equation}
\noindent
where  D$_{\rm L}$ is the luminosity distance to the source, and $\nu_{20 \mu m}$ is the frequency at $20$ \um. 

The radio luminosity at $5$ GHz, $L_{\rm 5GHz}$, is estimated as

\begin{equation}
L_{\rm 5GHz} = 4 \pi D_{L}^{2} \nu_{\rm 5GHz} \mbox{F}_{\rm 5GHz} (1+z)^{(-1-\alpha_{5})}
\label{eq:L5}
\end{equation}
\noindent
\citep{Sikora2007}, where $\nu_{\rm 5GHz}$ is the frequency at $5$ GHz, $z$ is the redshift, and $\alpha_{5} = 0.8$ for the K-correction.  We used the nuclear fluxes at $5$ GHz. Although the total radio emission (i.e., including extended lobes, hot spots, and jets) would provide a more isotropic measurement at radio wavelengths, the only effect would be a larger offset in the radio-loudness values between RL and RQ AGN.

Finally, the radio-loudness, $R_{20}$, is estimated as

\begin{equation}\label{eq:R20}
R_{20} = 2.99\times10^{3} \frac{L_{\rm 5GHz}}{L_{20\mu m}}
\end{equation}

Table \ref{tab:table2} shows the MIR from the literature, the estimated $L_{20\mu m}$, $L_{\rm 5GHz}$,  and $R_{20}$ for each AGN. Our sample covers a range of four orders of magnitude in $\log_{10} (L_{20\mu m} [\rm{erg~s^{-1}}]) = [41.6,45.4]$, and four orders of magnitude in $\log_{10} R_{20}= [-2.4,1.5]$. Figure \ref{fig:fig2} shows the trends between the $89$ \um\ intrinsic polarization with the $20$ \um\ luminosity, and radio-loudness, $R_{20}$.

The most interesting result is that the thermal polarization by means of magnetically aligned dust grains increases with the nuclear radio-loudness, $R_{20}$. Despite the small sample, the polarized and unpolarized AGN are well-defined at $\log_{10} (R_{20})>0.5$ and $\log_{10} (R_{20})< -1$, respectively. The RL AGN are within the range of $P_{\rm 89\mu m}^{\rm int} = [5,11]$\% (excluding NGC1275, Section \ref{subsec:DeltaPA}).  We find no clear correlations between the intrinsic thermal polarization with the radio and MIR luminosities. We show the first evidence of an intrinsic difference in the $89$ \um\ nuclear polarization between RL and RQ AGN.


\section{Discussion}\label{sec:DIS}


\subsection{Origin of the thermal polarization}\label{subsec:ThermalPol}

In this section, we show that our observations trace the B-field orientation by means of the thermal emission of magnetically aligned dust grains. As a quantitative analysis, we estimate the expected thermal polarized emission based on measurements of the absorptive polarization from magnetically aligned dust grains. Centaurus A has a nuclear polarization of $1.97\pm0.01$\% with a PA$_{\rm E}=134.09\pm0.16^{\circ}$ within an aperture of $8$\arcsec\ at $2.2$ \um\, and a visual extinction of A$_{\rm v}=16$ mag.  \citep{Packham1996}. The optical depth at K-band (2.0 \um) can be estimated such as $\tau_{\rm K} = 0.09A_{\rm v} = 1.44$ \citep{Jones1989}. Using the typical extinction curve\footnote{Synthetic extinction curve can be found at \url{https://www.astro.princeton.edu/~draine/dust/dustmix.html}} of the Milky Way with an extinction factor of $R_{\rm v} = 3.1$ \citep{Weingartner2001}, we estimate the optical depth at $89$ \um\ to be $\tau_{89} = 1.18\times10^{-2}\tau_{\rm K} = 1.69\times10^{-2}$. This result indicates that the dusty structure associated with the visual extinction to the core is optically thin at $89$ \um. Under the optically thin condition, the polarized thermal emission can be estimated as $P_{89}^{\rm em} = -P_{89}^{\rm abs}/\tau_{89}$. The negative sign indicates the $90^{\circ}$ change from absorptive to emissive polarization \citep{Hildebrand1988,Aitken2004}. The $90^{\circ}$ angular change is satisfied from $2.2$ \um, PA$_{\rm E}=134.09\pm0.16^{\circ}$, to $89$ \um, $PA_{\rm E}=45\pm4^{\circ}$, where we estimate an angular difference of $|\Delta\theta|=89\pm4^{\circ}$. As the extinction curve is representative of the absorptive polarization, the scaled measured $2.2$ \um\ polarization of $1.97\pm0.01$\% at $89$ \um\ is estimated to be $P_{89}^{\rm abs} \sim 0.02$\%. Finally, we estimate an expected emissive polarization of $P_{89}^{\rm em} \sim 1.18$\% at 89 \um. This result is in excellent agreement with the measured polarization of $1.5\pm0.2$\% at 89 \um\ (Table \ref{tab:table3}). Similar results were also found for NGC~1068 \citep{ELR2020}, NGC~1097 \citep{ELR2021c}, and using a detailed total and polarized SED fitting for Cygnus~A \citep{ELR2018b}.  In addition, we show in Section \ref{subsec:pSED} that the polarized synchrotron emission is negligible at FIR wavelengths for the polarized AGN. Thus, we conclude that the 89 \um\ polarization measurements of AGN arise from the thermal emission of magnetically aligned dust grains. The measured $PA_{E}$ (E-vector in Table \ref{tab:table3}) of polarization can be rotated by $90^{\circ}$ to show the inferred B-field orientation.

\subsection{Dust properties from FIR polarization}\label{subsec:DIS_dust}

Because the beam sizes of our $89$ \um\ observations are $>130$ pc (Table \ref{tab:table4}) and we mostly have a single wavelength observation per object (Table \ref{tab:table3}), it is very challenging to analyze the dust grain properties. We use the resolved, $\sim5$ pc, observations of 30 Doradus, a star-forming region in the Large Magellanic Cloud, as an example of thermal polarization emission properties in an external galaxy. Using $89$ \um\ polarimetric observations with HAWC+, \citet{Tram2021} measured polarization fractions up to $15$\% in 30 Doradus. The highest polarization arises from dust grains located at the lowest column density with dust temperatures $\le 37$ K around the star-forming regions. These authors argued that this polarization could be explained under the radiative alignment torques theory \citep[RAT][]{Andersson2015,HL2016}, where the polarization fraction depends on the dust grain alignment size. Specifically, the dust grain alignment size depends on the gas volume density, gas temperature, and radiation field in the ISM. In 30 Doradus, typical grain sizes of $0.01-0.82$ \um\ could explain the $15$\% polarization fraction. Despite the physical differences of 30 Doradus and at $\le130$ pc around AGN, these results show that high levels of polarized thermal emission can be observed under extreme conditions. In addition, these levels of polarization can be explained with nominal dust grains sizes and compositions. Note that for higher polarization fractions, as those up to $25$\% measured by \textit{Planck} \citep{PlanckIntXIX2015,PlanckXII2020}, other dust grain compositions (e.g., `astrodust') may be required \citep{Draine2021,Hensley2021}. 

Based on the results of emissive polarization in galaxies using FIR wavelengths, we conclude that typical dust grain sizes and composition in the ISM can explain our measured high polarization fractions of AGN. Note that `core' refers to spatial scales smaller than the beam size of our observations, i.e., from $<130$ pc to $<18.19$ kpc. Our results suggest that the FIR polarimetric observation may trace an ordered B-field within a dusty structure with low dust temperature and low surface brightness around the active nuclei. 


\subsection{Polarized emitting source}\label{subsec:Source}

We estimated the effective radius of the dusty structure emitting at 89 \um~assuming radiative equilibrium to be $r_{\rm{d,tk}}$=$[5,79]$ pc and $r_{\rm{d,th}}$=$[146,5802]$ pc for optically thick and optically thin dust, respectively (Section \ref{subsec:TIR}). We interpret $r_{\rm{d,tk}}$ as a lower-limit radius of the dusty structure surrounding the AGN. The $r_{\rm{d,th}}$ is substantially larger than $r_{\rm{d,tk}}$ because the AGN radiation can travel to larger distances through the dusty medium without being totally absorbed. Note that $r_{\rm{d,th}}$ is larger that, or similar to, the physical scale of the beam size of our observations at $89$ \um~for most of our AGN. However, the full range of distances up to the maximum of $r_{\rm{d,th}}$ cannot produce the polarization levels measured in RL AGN. We discuss the origin of the highly polarized dusty regions in RL AGN and the estimations of the effective radius.

Based on FIR polarimetric observations of resolved galaxies, spiral galaxies have a median $154$ \um~polarization fraction of $3.3\pm0.9$\% \citep{ELR2022b}. The lowest polarization arises from regions with high turbulence and/or tangled B-fields in the interstellar medium (ISM) due to galaxy interaction and star formation activity. These physical conditions generate an isotropic turbulent B-field and/or tangled B-fields along the line-of-sight, resulting in a net null polarization within the beam (typically $\sim300$ pc diameter) of the observations. The highest polarization mainly arises from low density, low surface brightness, low column density, low dust temperature, and low gas turbulence regions of the ISM. These conditions increase the polarization efficiency of dust grains with the local B-field in the host galaxies without invoking other physical conditions than typical dust parameters in the ISM. For both face-on and edge-on spiral galaxies without an (or a weak) AGN, the cores (beam size $\le1$ kpc) are measured to be unpolarized at FIR wavelengths \citep{Borlaff2021,ELR2022b}. The median $89$ \um~polarization fraction across the resolved $\sim3$ kpc dust lane of Centaurus A is $\sim4$\% \citep{ELR2021}. These results show individual polarization measurements with spatial resolutions from $90$ pc to $1$ kpc across the host galaxies. Note that RL AGN are typically hosted by early-type galaxies \citep{Kauffmann2003}, which may have different extended properties to those from spiral galaxies compared above. In addition, inactive galaxies have unpolarized cores at FIR wavelengths. These results suggest that, most likely, the high polarization in RL AGN may not arise from physical structures associated with the host galaxy below the beam size of our observations.

The integrated FIR polarization of spiral galaxies is $<1$\% \citep[table 4 and figure 3 by][]{ELR2022b}. This estimation was performed for the entire host galaxy covering spatial scales of $2-10$ kpc. This result rules out that the high polarization in unresolved RL AGN (i.e., Cygnus~A, PKS1345+12) may be attributed to the integrated polarization arising from the entire host galaxy. 

Knowing that the entire galaxy and physical structures within the host galaxy cannot produce high polarization, and using Centaurus~A as the example due to its proximity, we suggest that the polarized source in the core of RL AGN should arise from a structure $r_{\rm{d,th}}>5$ pc and $<130$ pc in radius directly radiated by the AGN.

The estimations of $r_{\rm{d,tk}}$ and $r_{\rm{d,th}}$ assume that the RL AGN solely heats the polarized dust.  In reality, the total flux at $89$ \um~may arise from a combination of dust heated by the AGN, dust heated by star formation around the AGN (i.e., starburst ring), and/or dust in the ISM of the host galaxy. However, not all these physical components can produce the high levels of FIR polarization measured in RL AGN. As the RQ AGN are unpolarized, we cannot constrain the physical sizes of the polarized structure, and the assumption that the RQ AGN solely heats the polarized dust may not apply to these objects. We explore these scenarios to explain the $5-11$\% polarization at $89$ \um~in RL AGN.

\cite{Tadhunter2007} measured that the AGN radiation, traced by [OIII] $\lambda5007$ emission, is the primary heating mechanism at $24$ \um~and $70$ \um. The scatter measured in [O III] and $70$ \um~may arise from starburst activity heating the cool dust at the lower-end of luminosity ranges, $\log_{10}(L_{\rm{70\mu m}} [\rm{erg~s^{-1}}]) < 42.6$, but only for $20-30$\% of objects. Star formation activity decreases the  FIR polarization due to tangled B-fields and/or increased turbulence in the ISM. Galactic shocks, as measured in the intersection of the galactic bars and starburst rings of NGC 1097 \citep{ELR2021c}, can compress the B-fields in the dense and cold ISM. Although a constant B-field of polarization parallel to the starburst ring would be measured for an edge-on view of the starburst ring, the $89$ \um~polarization is measured to be $\le3$\%. 

We estimated that the polarized synchrotron emission is negligible at $89$ \um~(Sections \ref{subsec:NTIR} and \ref{subsec:pSED}), and that the B-field orientation is highly offset from the radio jet axis in RL AGN (Section \ref{subsec:DeltaPA}). The radio jets in our RL AGN sample are mostly perpendicular to the galaxy disks. The radio jet in NGC1068 interacts with a giant molecular cloud at $\sim20$ pc North from the AGN producing high levels of polarization: $\sim7$\% at $10$ \um~\citep{ELR2016}, $5$\% at $860$ \um~\citep{ELR2020b}, and up to $11$\% at $860$ \um~in several knots within the central $100$ pc. However, the polarization mechanisms differ from those at $89$ \um. At $10$ \um, the polarization arises from dichroic absorption of a polarized region through the dust lane, while synchrotron polarization dominates at $860$ \um. In addition, the $89$ \um~polarization of the core and the bar of NGC~1068 is $<2$\% \citep{ELR2020}. All these results rule out the possibility that the polarization in RL AGN arises from dust heated by jet interaction in the ISM and star formation activity.


\subsection{Radio jet axis vs. inferred B-field orientation}\label{subsec:DeltaPA}

\begin{figure*}[ht!]
\includegraphics[scale=0.64]{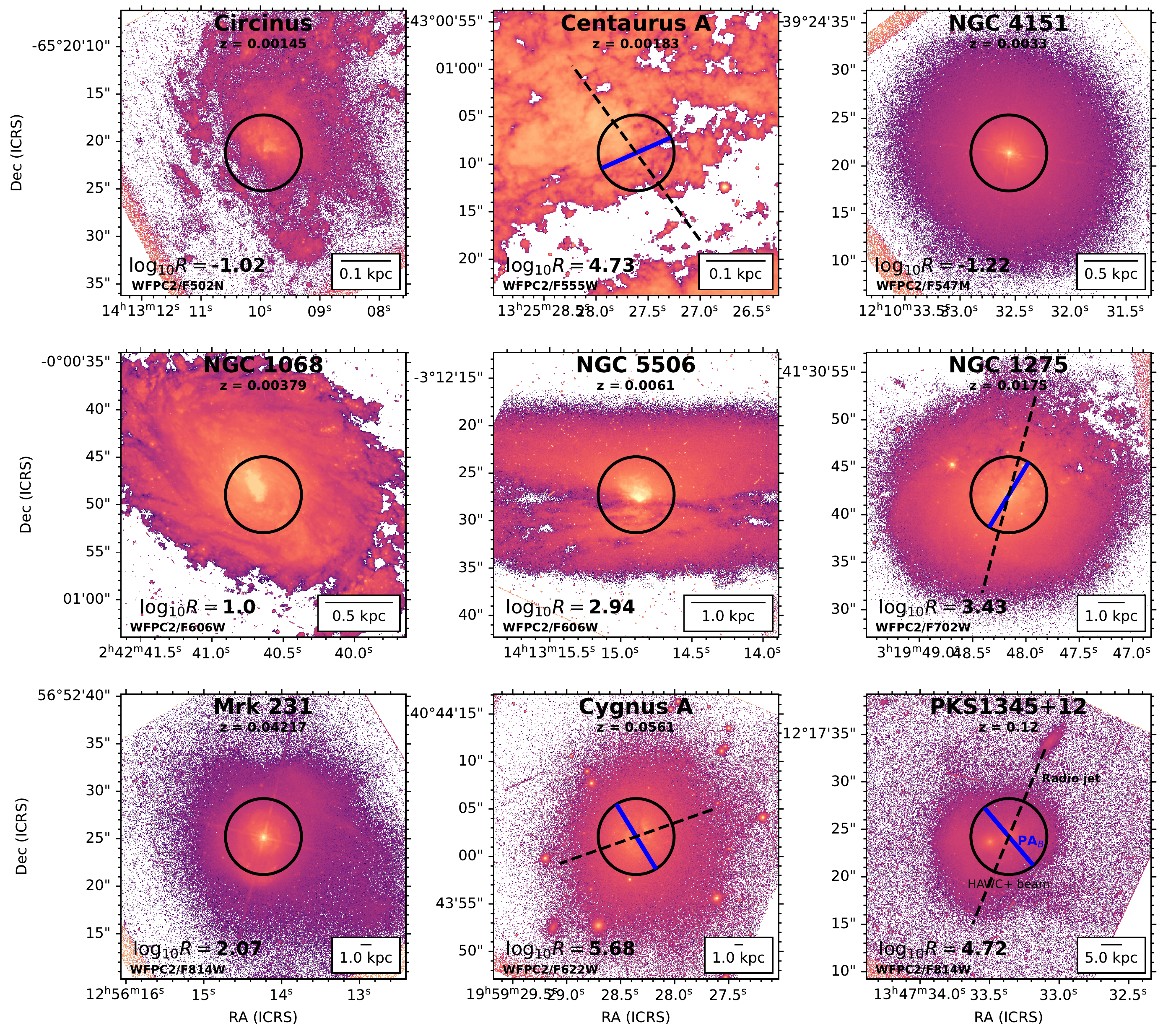}
\caption{$89$ \um~B-field and $5$ GHz radio jet orientations for the galaxies in our sample. The WFPC2\textit{HST} images (color scale) with their associated filters in the bottom left are shown. The B-field orientation (blue solid line) measured within the HAWC+ beam (black circle) and the radio jet axis orientation (black dashed line) are shown. Galaxies are ordered by their redshift, $z$, shown in the top center of each panel. A physical scale is shown in the bottom right of each panel.
 \label{fig:fig3}}
\end{figure*}

For the polarized objects of our sample, we compute the angular difference between the measured B-field orientation at $89$ \um\ and the nuclear jet axis, i.e., jet-disk (Table \ref{tab:table5}). Figure \ref{fig:fig3} shows our measured B-field orientation at $89$ \um\ and the $5$ GHz radio jet orientation over the WFPC2/\textit{HST} images. We find that $75$\% (3 out of 4 objects) of the polarized cores have a jet-disk angular difference in the range of $\Delta\theta = 64-84^{\circ}$, with a median of $\langle \Delta\theta \rangle = 65\pm26^{\circ}$. The exception is  NGC~1275 with an angular difference of $19\pm2^{\circ}$. Despite the small sample, we find several interesting results.

\begin{deluxetable}{lcccp{2.8cm}}[ht!]
\tablecaption{Jet axis and magnetic field orientation at 89 \um.\label{tab:table5}}
\tablecolumns{7}
\tablewidth{0pt}
\tablehead{\colhead{Object}& \colhead{PA$_{\rm jet}^{1}$} & \colhead{PA$_{\rm B}$}	&	\colhead{$\Delta\theta$}	&	\colhead{References}	 \\ 
&	\colhead{($^{\circ}$)}  & \colhead{($^{\circ}$)}	&	\colhead{($^{\circ}$)}	 
}
\startdata
Centaurus~A	&	$51$		&	$135\pm4$	&	$84\pm4$		&	$^{1}$\citet{Israel1998} \\
Cygnus~A		&	$105$	&	$39\pm7$		&	$66\pm7$		&	$^{1}$\citet{Sorathia1996} \\
NGC~1275	&	$160$	&	$141\pm2$	&	$19\pm2$		&	$^{1}$\citet{Pedlar1990} \\
PKS1345+125	&	$157$	&	$41\pm8$		&	$64\pm5$	&	$^{1}$\citet{Lister2003}
\enddata
\end{deluxetable}

The inferred B-field orientation is highly offset with respect to the jet axis, but not exactly at $90^{\circ}$.  Centaurus~A is known to have a warped disk from $2$ to $6500$ pc \citep{Quillen2010}. Using our estimated beam size of $8$\arcsec\ (radius of $4\arcsec=64$ pc) at $89$ \um, we find that the position angle of the velocity component is $124.8^{\circ}$ at a radius of $3.99\arcsec$ from the core \citep[table 1 by][]{Quillen2010}. The associated ring of matter at a radius of $64$ pc around the center of Centaurus~A is almost parallel, $|135^{\circ}-124.8^{\circ}|=10.2^{\circ}$, to the measured B-field orientation. This result indicates that the measured B-field may be spatially associated with the gas flows at $64$ pc from the core of Centaurus A. Using $2$ \um\ polarimetric observation with NICMOS/\textit{HST}, \citet{Tadhunter2000} found that the ionization cone of Cygnus~A has only polarized along one of the edges up to $1.3$ kpc from the core. The $2$ \um\ polarization is thought to arise from dust scattering in the ionization cone. These authors argued that the observed polarization is due to intrinsic anisotropic radiation caused by a warped disk. A disk of $80-100$ pc in radius roughly perpendicular to the radio jet axis was observed using the Very Long Baseline Array (VLBA) in HI absorption \citep{Struve2010a}. These results indicate that the measured B-field may be spatially associated with the warped disk at $24.7$ pc from the core of Cygnus A. The radio jet of PKS1345+125 is thought to be precessing with a cone half-angle of $23^{\circ}$ \citep{Lister2003}. Indeed, misaligned disks have been measured in the nuclear structures of AGN \citep[i.e.][]{LE2010}. 

These results show that the most plausible interpretation of our measurements is that the measured B-field orientation at $89$ \um\  may be associated with a toroidal B-field along the equatorial axis of the AGN. The angular difference between the equational and jet axes, $\Delta\theta$, can be explained by intrinsic anisotropies of the magnetized structure and jet axis (i.e., warped disk, precessing jets, MHD wind).

The exception is NGC~1275, where the inferred B-field orientation seems almost parallel, $\Delta\theta=19\pm2^{\circ}$, with the jet axis. Our $89$ \um\ dust continuum emission observations show extended emission at an $11\sigma$ detection with a PA $\sim125^{\circ}$ out to a radius of $\sim12$ kpc (Appendix \ref{app:NGC1275}). NGC~1275 shows thermal emission features extended at PAs of $\sim120^{\circ}$ and $\sim75^{\circ}$ using $70-160$ \um\ \textit{Herschel} observations with beam sizes of $5-12$\arcsec\ \citep{Mittal2012}. The thermal emission at 450 \um\ with a beam size of $9.1\arcsec\times8.8\arcsec$ using the James Clerk Maxwell Telescope (JCMT) also shows an extended emission at $\sim115^{\circ}$ up to a radius of $\sim20$ kpc \citep{Irwin2001}.  These observations indicate that a large amount of dust is present in this galaxy and the surrounded Perseus cluster \citep[$\sim6\times10^{7}$M$_{\odot}$ with $T_{\rm d}\sim30$ K and $\beta=1.3$;][]{Irwin2001}. Given the low angular resolution of our observations, it is unclear how to spatially identify a single structure with the measured B-field orientation.  We speculate that a physical structure associated with the extended dusty emission along the NW region at a PA of $\sim125^{\circ}$ may contain a dominant ordered B-field. This magnetized structure may be associated with one of the H$_{\alpha}$ \citep{Caulet1992,Conselice2001} and CO filaments \citep{Nagai2019}. In addition, the contribution of the synchrotron emission from these filaments at FIR wavelengths is unknown. Further high-angular resolution ($<1$\arcsec) IR total and polarimetric observations are required to identify the associated physical structure with the inferred B-field orientation at $89$ \um. The $89$ \um~polarimetric observations of NGC~1275 are inconclusive and will not be used for the physical interpretation discussed in Section \ref{subsec:RQvsRL}.


\subsection{The strength and orientation of the B-fields in the polarized dusty structure}\label{subsec:BfieldFIR}

At $10-100$ pc scales, we have shown that the nuclear thermal dust polarization at 89 \um\  arises from magnetically aligned dust grains (Sections \ref{subsec:pSED} and \ref{subsec:ThermalPol}), and increases with the nuclear radio-loudness (Section \ref{sec:R}). In addition, we show the presence of a $5-130$ pc-scale ordered toroidal B-field in RL AGN but not in RQ AGN (Section \ref{subsec:DeltaPA}). 

At $1-10$ pc scales, the inner wall of the dusty torus at scales of $1-2$ pc is dominated by an ordered toroidal B-field in the RL AGN IC~5063 \citep{ELR2013} and RQ AGN NGC~1068 \citep{ELR2015}. The B-field orientation was measured to be parallel to the equatorial axis of the dusty torus. The B-field strength in the RL AGN IC~5063 \citep[$\sim41$ mG;][]{ELR2013} was measured to be a factor ten larger than for the RQ AGN NGC~1068 \citep[$\sim3$ mG;][]{ELR2020b} at $1$ pc radius from the active nuclei. These results are based on the characterization of hot ($\sim1300$ K) dust at scales of $1$ pc from the central engine. ALMA polarimetric observations at $860$ \um\ found a magnetized dusty torus in NGC~1068 \citep{ELR2020b}. The polarization arises from magnetically aligned dust grains, where the inferred B-field orientation was measured to be parallel to the equatorial axis within the $3-8$ pc eastern side of the dusty torus, i.e., toroidal B-field. In the same region of the torus, the B-field strength was estimated to be $0.67^{+0.94}_{-0.31}$ mG. These results provide the B-field configuration and physical mechanisms at the inner boundaries ($\sim$pc-scales) of our findings. 

Following these results and assuming conservation of the magnetic energy (i.e., `flux-freezing') and perfect conductivity (i.e., without diffusion), the toroidal B-field strength decreases with distance from the core as $B\propto r^{-1}$ \citep[for a toroidal B-field $B_{\phi} \sim \rho r$;][]{Begelman1984}.  For RL AGN, we take a $B=41$ mG at $1$ pc \citep{ELR2013}, and estimate a B-field strength of $8-0.3$ mG at $5-130$ pc. For RQ AGN, we take the measured B-field strength of $\sim0.67$ mG at $5$ pc in NGC~1068 \citep{ELR2020b}, and estimate a B-field strength of $28$ $\mu$G at $130$ pc from the core. Thus, a stronger B-field associated with the gas flow in RL AGN may be present at larger distances from the core than in RQ AGN.

We measured a median offset of $65\pm26^{\circ}$ between the inferred B-field orientation and the radio jet for RL AGN (Section \ref{subsec:DeltaPA}). Recent MHD simulations have shown that the infalling gas is not rotationally supported and moves radially inwards, which may generate a highly asymmetric disk at scales of $10-100$ pc \citep{Takasao2021}. The accreting material from large scales ($\sim100$ pc) to the accretion flow at scales of $<10$ pc may produce a warped disk \citep[i.e.,][]{LE2010}. Furthermore,  Centaurus A \citep{Graham1979,Struve2010b}, Cygnus A \citep{Struve2010a}, and PKS 1345+125 \citep{Gilmore1986,Lister2003} show signs of mergers, which may cause a warped disk with a warped B-field orientation \citep{ELR2021}, and may also trigger episodes of hot accretion \citep{SB2013}. Any of these mechanisms (i.e., disk instabilities, mergers) may reproduce the observed offsets between the inferred B-field orientation and the radio jet axis in the core of RL AGN (Section \ref{subsec:DeltaPA}).

\subsection{Physical interpretations of the RL and RQ AGN dichotomy}\label{subsec:RQvsRL}

Our results may be associated with a physical structure in the interface between the sphere of influence of the SMBH and jet production at sub-pc scales and the host galaxy at kpc scales. We need to connect our measurements with the physical mechanisms and structures associated with generating B-fields at sub-pc and kpc scales.

At kpc-scales, the B-field in the dense, N$_{\rm HI+H2}=[10^{20},10^{23}]$ cm$^{-2}$, and cold, $T_{\rm d}=[19,48]$ K, ISM of nearby galaxies hosting AGN have been recently traced using FIR polarimetric observations \citep[e.g.,][]{ELR2022b}. For spiral galaxies, a kpc-scale B-field cospatial with the optical spiral arms has been measured for NGC~1068 \citep{ELR2020} and NGC~1097 \citep{Beck2005}. NGC~1068 and NGC~1097 have an inner bar of a few kpcs in length, where the B-field is compressed with an orientation parallel to the long axis of the bar. Within the central $1$ kpc of NGC~1097, the B-field in the dense ISM (traced in FIR wavelengths) is compressed by a shock driven by the galactic bar dynamics around the starburst ring. The B-field in the diffuse ISM (traced in radio wavelengths) is thought to dominate the infalling material from kpc-scale to $100$ pc-scale toward the AGN \citep{Beck2005,ELR2021c}.  Specifically, the mixing of the radio emission from relativistic electrons and the diffuse ionized ISM responds to shear flows, which drag the diffuse gas via an MHD dynamo toward the active nucleus. The B-field strength is estimated to be $60$ $\mu$G at $\sim500$ pc from the AGN in NGC~1097 \citep{Beck2005}. These observations show that an MHD dynamo with a kpc-scale spiral B-field allows the flow of gas at kpc-scales from the galaxy hosting AGN to move inwards toward the core up to scales of $\sim100$ pc. These results provide the B-field configuration and physical mechanism at the outer boundaries of our findings, i.e., an ordered spiral B-field on an MHD dynamo feeding gas from kpc to $\sim100$ pc scales toward the SMBH.

At sub-pc scales, we describe two main scenarios. On the one hand, magnetically elevated accretion disks provide the physical structure for the transfer of angular momentum and the production of jets and outflows by the combination of MRI and accretion disk dynamo \citep{Lubow1994, GO2012, Begelman2017}. A weak magnetic field can grow in time to significantly influence the dynamics of the accretion disk. Specifically, the jet power depends on the location of the emitting region along the equatorial plane and the strength and geometry of the B-field at that location. For a sample of 229 Blazars, \citet{Ghisellini2014} found that the dominant emitting region along the equatorial plane of the AGN is located at smaller scales than the radius of the broad line regions (BLR) for 85\% of the sources and between the BLR and the inner radius of the dusty torus for 15\% of the sources. These authors stated that if the emitting region is located at larger distances ($\ge$ pc-scales), where the emitting structure is unaffected by external radiation, the jet power will increase. Thus, a strong and ordered toroidal B-field at scales of $1-10$ pc will enhance the accretion flow and, therefore, the jet power. 

Under this first scenario, an MHD accretion flow at large radii (up to $\sim100$ pc) may be the dominant mechanism for the RQ and RL dichotomy. Our results indicate that the stronger the jet power, the larger distances from the SMBH the B-field may affect the accretion flow. We showed that a $5-130$ pc-scale ordered toroidal B-field is present in RL AGN, which we suggest may be the outer layers of the accretion flow of the AGN and the cause of the radio-loudness. Whether the polarized thermal emission structure spatially corresponds with the outer edge of the dusty torus ($<50$ pc) or a larger ($50-100$ pc) circumnuclear dusty environment may be an issue of semantics.

This first scenario has several caveats. The dynamics of the gas flow and how the MHD dynamo from large scales interpolates to $<130$ pc, making the dynamo more efficient, is unknown. In addition, for the magnetically elevated disk to be efficient at these distances, the energy from organized vertical B-fields has to become comparable with the turbulent and thermal energy, and the MRI has to operate in the molecular and dusty disk at $<130$ pc. \citet{SB2013} proposed a scenario where hot accretion events can trap magnetic flux in the disk that then is released due to the loss of angular moment from the SMBH.  Further model development of the effect of MRI in a rapidly rotating flow is required to test this scenario.

On the other hand, the radio jet generates a strong and large-scale toroidal B-field with a current pointing outwards \citep[i.e.][]{Begelman1984}. It has also been suggested that under certain conditions, some of the rotational power of the spinning black hole can be transmitted to the disc to form an MHD wind that drives away most of the mass \citep{BG2022}. This scenario implies that there is an infall of matter at large distances (i.e., Bondi radius) from the SMBH, sufficient to retain the magnetic flux to extract the energy from the SMBH. The median of gas mass in the tori using a sample of 19 Seyfert galaxies was estimated to be $\sim6\times10^{5}$ M$_{\odot}$ ($\sim10^{39}$ g) \citep{GB2021}. We estimated a B-field strength of $8-0.3$ mG at a radius of $5-130$ pc from the SMBH in RL AGN. These results are larger than the minimum requirements described in \citet{BG2022}, which may satisfy this condition. If the infalling gas has a specific angular momentum $>GM_{\rm BH} /\sigma$, a toroidal B-field could be generated within a disk over the region observed by HAWC+/SOFIA. This scenario also requires a large-scale and toroidal B-field in the presence of a powerful MHD wind, but this will likely arise within the Bondi radius. Under this second scenario, the fact that RL AGN already have powerful jets may imply that the radio-loudness causes the measured B-field from our observations.

This second scenario has several caveats. The B-field morphology and strength at pc-scales depend on the current being unscreened. It is unknown how far away the outward current, generated by the jet power, closes when the jet propagates into the ISM. In addition, the accretion disk may be highly turbulent, and the B-field outside the jet may diffuse rapidly into a free-free configuration \citep[i.e.][]{Begelman1984}. As an example, the resolved B-fields in the dusty torus of NGC 1068 using $860$ \um\ polarimetric observations showed highly turbulent gas motions on the western side of the equatorial torus. Interestedly, the ordered and toroidal B-field in the dusty torus of NGC 1068 was only measured in the eastern region, which is associated with low turbulent gas motions in the disk \citep{ELR2021a}. Finally, this second scenario has been applied to M87, and SOFIA/HAWC+ observations of M87 are not available to further test this model.

The fact that there is an ordered toroidal B-field at scales of $5-130$ pc in RL AGN implies that the MHD models of accretion in SMBH and jet production may need to be connected with the MHD dynamo in the host galaxy. Our results show that the AGN dusty torus needs to be studied within an MHD framework.


\section{Conclusions}\label{sec:CON}

We presented the nuclear thermal dust polarization of a sample of $9$ AGN in the $53-214$ \um\ wavelength range using HAWC+/SOFIA. For all objects, the total flux nuclear SEDs show an IR bump at $100\pm20$ \um\ with a characteristic dust temperature of $34.8\pm7.8$ K. The only difference in the nuclear total SEDs resides in the synchrotron emission from the jet at radio wavelengths, which divides the sample in RL and RQ AGN. We find that this is not the case for the FIR polarized fluxes.

We measured that the cores of RL AGN are intrinsically polarized at $5-11$\% at $89$ \um, while RQ AGN are unpolarized ($<0.4$\%). We only corrected for the dilution of the measured polarization by the galaxy's dust lane in Centaurus A. The rest of RL AGN are point sources, and the measured polarization is estimated to be the intrinsic polarization. The measured polarization in RL AGN is detected even with the different spatial scales ($\sim0.1-18$ kpc) probed for different galaxies. Recent FIR polarimetric observations of nearby galaxies showed a median $154$ \um~polarization of $3.3\pm0.9$\% and $<1$\% within the resolved galaxy disks and their integrated polarization, respectively \citep{ELR2022b}. These results indicate that the FIR polarization of RL AGN is sensitive to a polarized dusty region below the spatial resolution of the observations ($\sim0.13-18.19$ kpc).  For RQ AGN, any potential polarization arises from magnetically aligned dust grains in the host galaxy within the angular resolution of the observations ($\sim0.15-6.94$ kpc), and the cores are unpolarized at a level of $\le0.4$\%. 

We computed the polarized SEDs of the RL AGN and found that 1) the synchrotron polarization is negligible at FIR wavelengths, and 2) the polarized FIR SEDs have the same shape as the IR bump in total flux. We showed that the measured $89$ \um~polarization arises from magnetically aligned dust grains from a dusty environment with a radius of $5-130$ pc. We estimated that the inferred B-field orientation is highly offset, $\langle \Delta\theta \rangle = 65\pm26^{\circ}$, with respect to the jet radio axis, but not exactly $90^{\circ}$. This result may be explained by intrinsic anisotropies between the magnetized dusty structure and the jet axis (i.e., warped disk, precession jets, MHD wind). 

We estimated the nuclear radio-loudness, $R_{20}=L_{5GHz}/L_{20\mu m}$. We found that the thermal polarization increases with increasing radio-loudness. Specifically, AGN are unpolarized at $\log_{10}(R_{20}) < -1$, and polarized at $\log_{10}(R_{20}) > 0.5$. We used the $20$ \um\ nuclear flux as a calorimeter of the big blue bump, which is independent of inclination effects, reddening, and/or beaming effects. This result shows a new observational difference in the RL and RQ dichotomy. We conclude that this result is due to the presence of a $5-130$ pc-scale ordered toroidal B-field cospatial with the equational axis of the dusty environment surrounding RL AGN. 

We discuss several potential scenarios based on the MHD accretion flow toward AGN and MHD winds. The former can be interpreted as the cause of radio-loudness, while the latter can be interpreted as the effect of radio-loudness. Under the first scenario, we suggest that the measured B-field is part of the accretion flow in RL AGN that ultimately supports the transfer of gas inwards and jet production. Under the second scenario, we suggest that the measured B-field is due to a current and the transport of angular momentum moving outwards generated by the jet power. An MHD wind is generated to extract the energy from the SMBH, and an infall of matter at large distances is required to retain magnetic flux in the disk. Our results show a new and previously unobserved difference between RL and RQ AGN. To distinguish between these two scenarios, further observations of several emission lines (i.e., OIV, NeIII, NeV, H2 S(3))  tracing the inflow and outflows of gas within the central $\le$ 100 pc of RL and RQ AGN are required. These observations are well-suited for using integral field units (IFU) at NIR and MIR wavelengths using the \textit{JWST}. In addition, ALMA polarimetric observations can provide a more extensive survey of RL and RQ AGN at $z=1-2$ to increase the statistics of these findings.


\begin{acknowledgments}
The authors thank the reviewer for their helpful comments, which improved the quality of the manuscript. Based on observations made with the NASA/DLR Stratospheric Observatory for Infrared Astronomy (SOFIA) under the 05\_0071, 08\_0012, and 07\_0032 Programs. SOFIA is jointly operated by the Universities Space Research Association, Inc. (USRA), under NASA contract NNA17BF53C, and the Deutsches SOFIA Institut (DSI) under DLR contract 50 OK 0901 to the University of Stuttgart. 
\end{acknowledgments}


\appendix

\section{PSF-fitting results of Centaurus A}\label{app:CenA}

\begin{figure*}[ht!]
\centering
\includegraphics[scale=0.35]{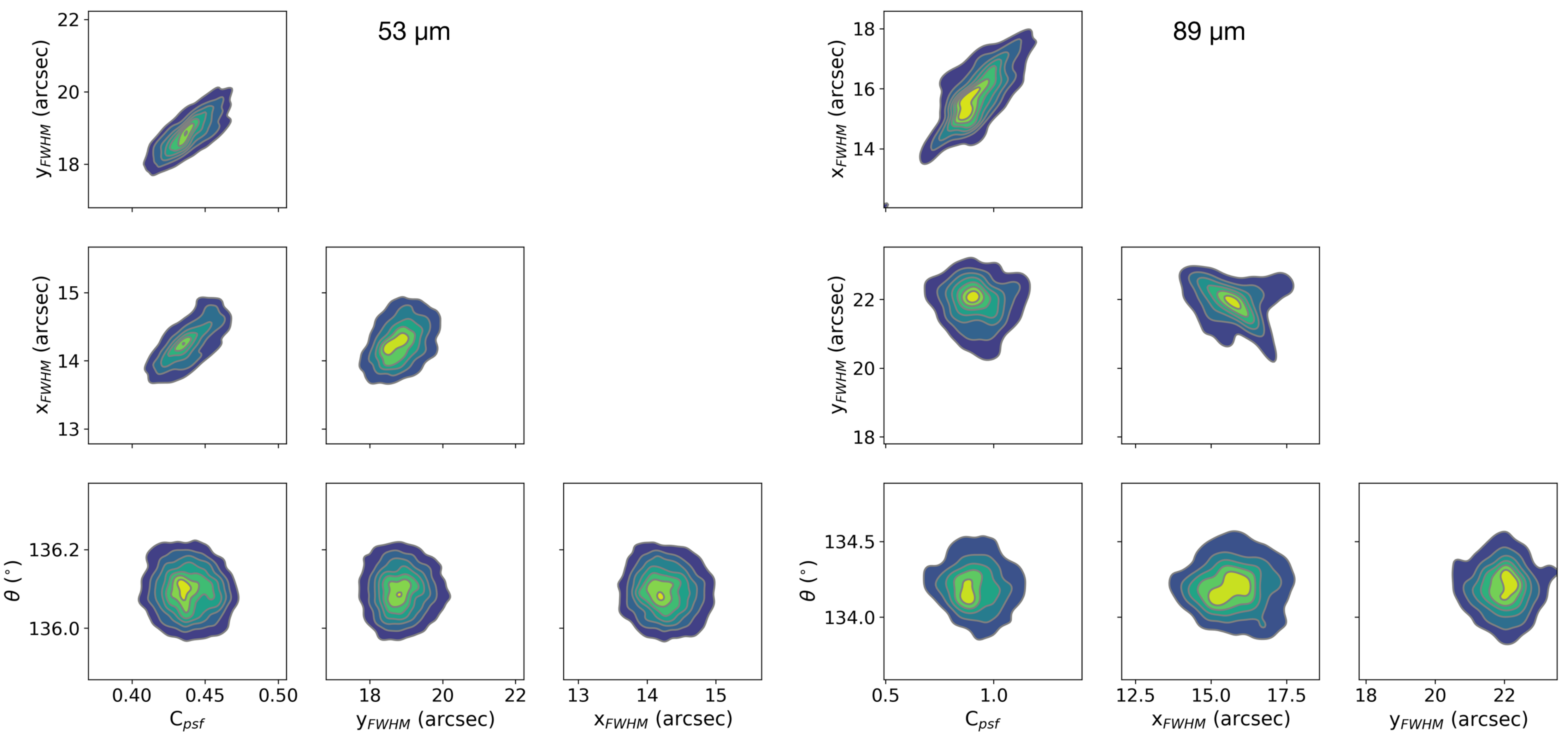}
\caption{Posterior distributions of the four free parameters shown in Table \ref{tabApp:table1} at $53$ (left) and $89$ (right) \um.
 \label{figApp:fig1}}
\end{figure*}

\begin{figure*}[ht!]
\centering
\includegraphics[scale=0.35]{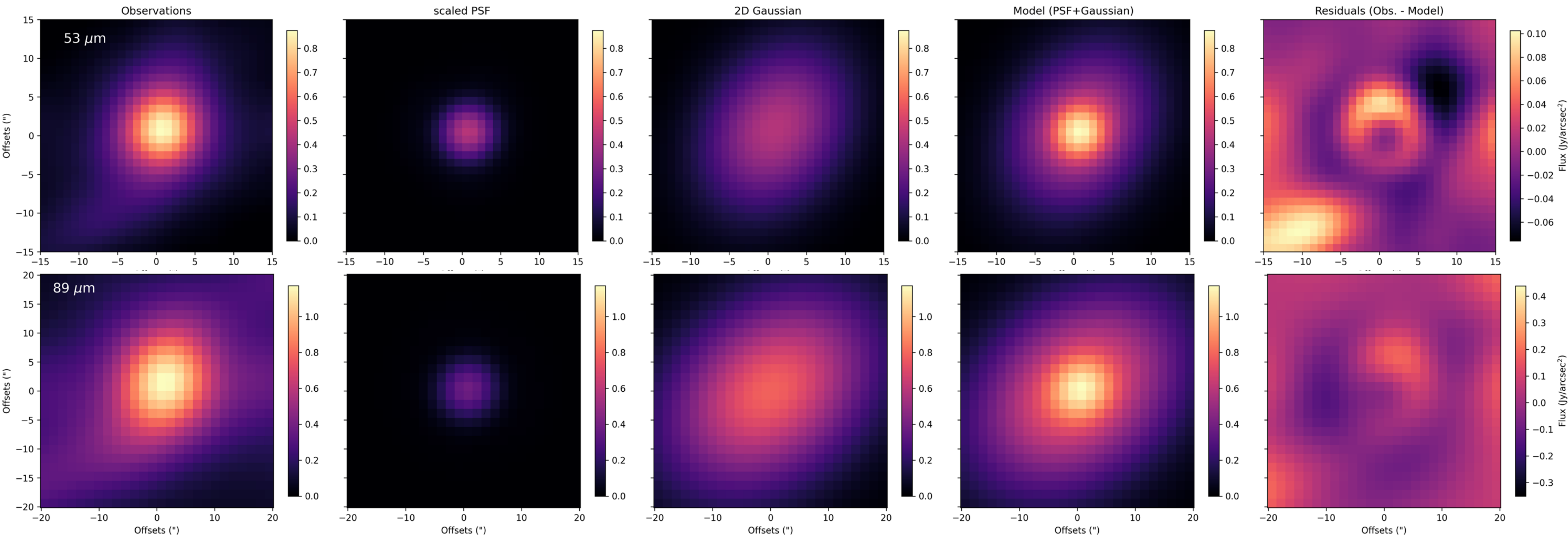}
\caption{Results of the PSF-fitting method. From left to right: Centaurus A observations, scaled PSF, 2D Gaussian, model (PSF+2D Gaussian), and residuals (observations - model) of the central $3\times$FWHM radius at $53$ (top) and $89$ (bottom) \um.
 \label{figApp:fig2}}
\end{figure*}

This section shows the results of the PSF-fitting routine for Centaurus A described in Section \ref{subsec:IP}. Table \ref{tabApp:table1} shows the best-inferred results of the four free parameters of the PSF-fitting method, and the measured fluxes for each component. Figure \ref{figApp:fig1} shows the posterior distributions of the four free parameters used in the PSF-fitting method at $53$ and $89$ \um. Figure \ref{figApp:fig2} shows the images of the observations, model components, and residuals for each wavelength.

\begin{deluxetable*}{cccccccccc}[ht!]
\tablecaption{Best fit parameters and modeled flux contributions of the PSF-fitting method described in Section \ref{subsec:IP}. (a) Wavelength of the HAWC+ observations. (b) $C_{PSF}$ is the peak flux of the PSF. (c,d) FWHM$_{a,b}$ are the FWHM of the short, $a$, and long, $b$, axes of the 2D Gaussian. (e) $\theta$ is the PA of the 2D Gaussian measured East from North. (f) F$_{T}$ is the measured total nuclear flux within the central beam. (g) F$_{PSF}^{M}$ is the measured nuclear PSF flux within the central beam. (h) F$_{gal}^{M}$ is the measured 2D Gaussian nuclear flux within the central beam. (i) F$_{T}^{M}$ is the measured nuclear total flux within the model. (j) and $\%PSF$ is the relative contribution of the PSF within the central beam. 
\label{tabApp:table1}}
\tablewidth{0pt}
\tablehead{\colhead{Wavelength}& \colhead{C$_{PSF}$} 	&	\colhead{FWHM$_{a}$}		&	 \colhead{FWHM$_{b}$}	&	\colhead{$\theta$} & \colhead{F$_{T}$} 	&	\colhead{F$_{PSF}^{M}$}		&	 \colhead{F$_{gal}^{M}$}	&	\colhead{F$_{T}^{M}$}	&	\colhead{\%PSF}	\\ 
\colhead{(\um)}  & \colhead{(Jy)}	& \colhead{(\arcsec)}	&	\colhead{(\arcsec)}	&	\colhead{($^{\circ}$)} &	\colhead{(Jy)}	& \colhead{(Jy)}		&	\colhead{(Jy)}	&	\colhead{(Jy)} &		\\
\colhead{(a)} & \colhead{(b)} & \colhead{(c)} & \colhead{(d)} & \colhead{(e)} & \colhead{(f)} & \colhead{(g)} & \colhead{(h)} & \colhead{(i)} & \colhead{(j)}
}
\startdata
53	&	$0.43^{+0.02}_{-0.03}$		&	$14.1^{+0.6}_{-0.6}$		&	$18.3^{+0.6}_{-0.6}$		&	$136^{+1}_{-1}$ &	$3.31\pm0.33$	&	$1.55\pm0.08$	&	$1.75\pm0.09$	&	$3.30\pm0.17$	&	$47\pm5$	\\
89	&	$0.93^{+0.25}_{-0.22}$		&	$15.8^{+2.0}_{-1.7}$		&	$21.8^{+1.6}_{-1.6}$		&	$134^{+1}_{-1}$ &	$10.8\pm0.11$	&	$3.48\pm0.17$	&	$7.51\pm0.38$	&	$11.0\pm0.55$	&	$32\pm3$
\enddata
\end{deluxetable*}

\section{Archival polarimetric measurements}\label{app:PolSED}

Table \ref{tab:table7} shows the polarimetric measurements from the literature used for the polarized SED shown in Figure \ref{fig:fig1}.

\begin{deluxetable*}{lccccl}[ht!]
\tablecaption{Polarimetric measurements from the literature. (a) Object name. (b) Central wavelength of the polarimetric observation. (c) Angular and spatial size of the aperture used for the photometric and polarimetric analysis. (d) Polarization fraction in \%. (e) Polarized flux in mJy. (f) References of the polarimetric observations.
\label{tab:table7}}
\tablecolumns{7}
\tablewidth{0pt}
\tablehead{\colhead{Object}& \colhead{Band} 	&	\colhead{Aperture}		&	\colhead{P$_{m}$}	&\colhead{PI}  &  \colhead{Reference}\\ 
					&	\colhead{(\um)}  &	\colhead{(\arcsec) / (kpc)}	& \colhead{(\%)}	& \colhead{(mJy)}	  &	\\
\colhead{(a)} & \colhead{(b)} & \colhead{(c)} & \colhead{(d)} & \colhead{(e)} & \colhead{(f)}
}
\startdata
Centaurus~A		&	800		&	$6$ / $0.10$					&	$0.89\pm0.46$	&	$\le0.07$			&	\citet{Packham1996,Hawarden1993} \\
				&	1100		&	$6$ / $0.10$					&	$0.12\pm0.20$	&	-				&	\citet{Packham1996} \\
				&	59958	&	$31\times10$ / $0.46\times0.16$	&	$0.21\pm0.01$	&	$14.70\pm1.10$ 	& \citet{Burns1983} \\
				&	214137	&	$3.6\times1.1$ / $0.06\times0.02$	&	$0.46\pm0.03$ &	$16.40\pm1.10$ 	& \citet{Burns1983} \\
Cygnus~A			&	2.2		&	$0.375$ / $0.42$				&	$10\pm1.5$	&	$0.49\pm0.04$		& \citet{Tadhunter2000} \\
				&	8.7		&	$0.38$ / $0.42$					&	$11\pm3$		&	$3.19\pm0.66$		& \citet{ELR2018b} \\
				&	11.6		&	$0.38$ / $0.42$					&	$12\pm3$		&	$5.40\pm2.08$		& \citet{ELR2018b} \\
				&	1400		&	$12$ / $13.32$					&	null			&	$<3.9$			&\citet{Ritacco2017} \\
				&	2000		&	$18.2$ / $20.20$				&	null			&	$<1.5$			& \citet{Ritacco2017} \\
PKS1345+125		&	20000	&	$1.42\times0.64$ sqmas / $3.2\times1.46$ pc$^{2}$				&	$<0.3$		&	$<9.9$			& \citet{Lister2018}	\\
				& 30000-180000	&	$1.4$ ($3.2$)		&	$<0.2$	&	$<2$		& \citet{Stanghellini1998}\\
\enddata
\end{deluxetable*}

\section{Total flux image of NGC~1275}\label{app:NGC1275}

This section shows the total intensity image of NGC~1275 analyzed in Section \ref{subsec:DeltaPA}.

\begin{figure}[ht!]
\centering
\includegraphics[scale=0.35]{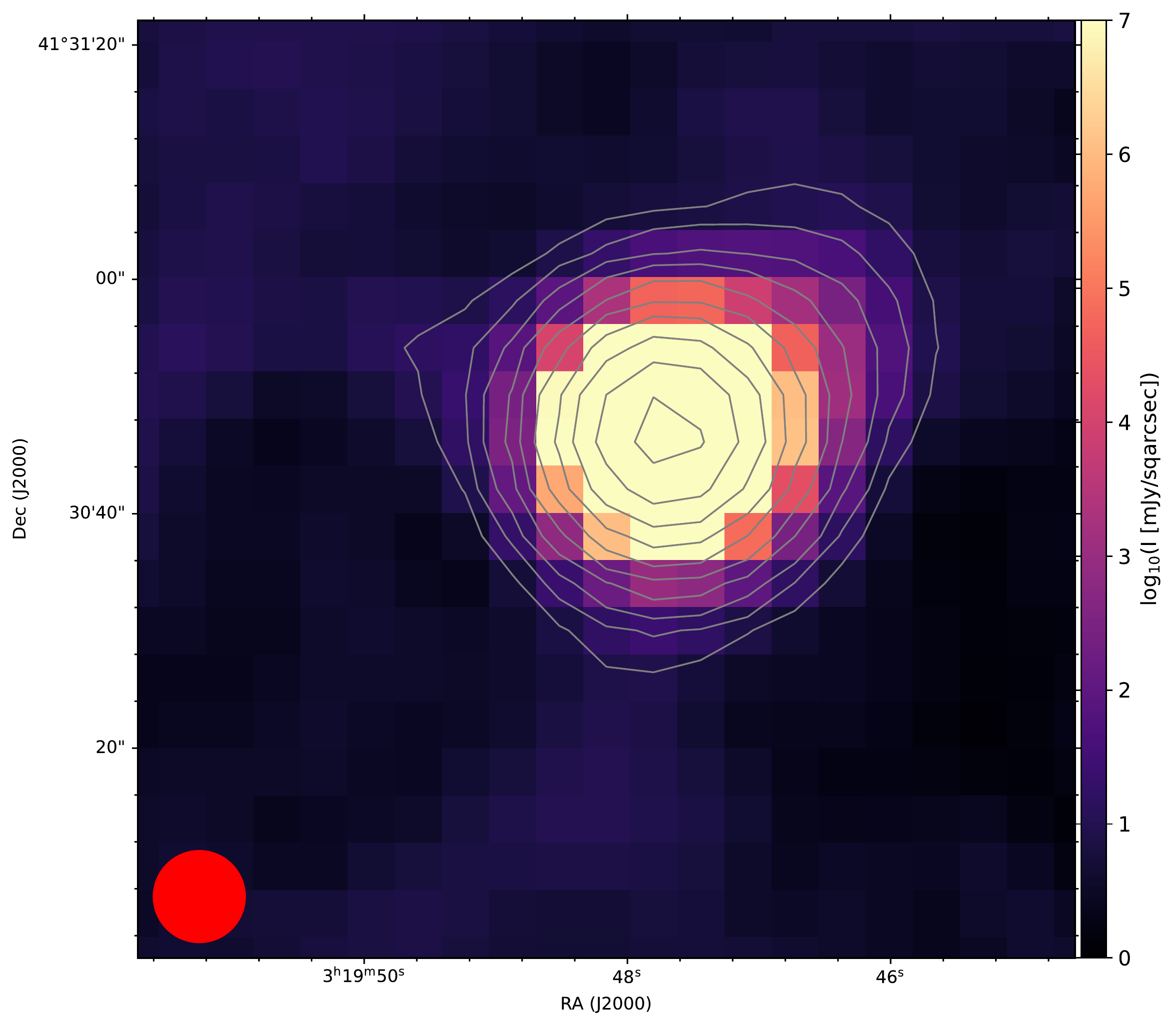}
\caption{Total flux image of NGC~1275 at $89$ \um. Contours start at $11\sigma$ and increase in steps of $\sigma2^{n}$, where $\sigma=0.094$ mJy/sqarcsec, and $n=3.5, 4.5, 5.5, \dots$. The beam size (red circle) is shown.
 \label{figApp:fig3}}
\end{figure}

%

\vspace{5mm}
\facilities{SOFIA (HAWC+)}


\software{aplpy \citep{aplpy},  
          astropy \citep{astropy}
          pymc3 \citet{pymc}}

\bibliography{references}{}
\bibliographystyle{aasjournal}



\end{document}